\documentclass[times,twocolumn,final]{elsarticle}

\usepackage{medima}
\usepackage{framed,multirow}

\usepackage{amssymb}
\usepackage{latexsym}

\usepackage{url}
\usepackage{xcolor}

\usepackage{hyperref}

\usepackage{amsmath} 

\definecolor{newcolor}{rgb}{.8,.349,.1}

\journal{Medical Image Analysis}

\begin{document}

\verso{Hanxiao Zhang \textit{et~al.}}

\begin{frontmatter}

\title{Faithful learning with sure data for lung nodule diagnosis}

\author[1]{Hanxiao \snm{Zhang}}

\author[2]{Liang \snm{Chen}}

\author[3]{Xiao \snm{Gu}}

\author[1]{Minghui \snm{Zhang}}

\author[1]{Yulei \snm{Qin}}

\author[2]{Feng \snm{Yao}}

\author[2]{Zhexin \snm{Wang}}

\author[1]{Yun \snm{Gu}\corref{cor1}}
\cortext[cor1]{Corresponding author: 
}
\ead{geron762@sjtu.edu.cn}
\author[1]{Guang-Zhong \snm{Yang}\corref{cor2}}
\cortext[cor2]{Corresponding author: 
}
\ead{gzyang@sjtu.edu.cn}

\address[1]{Institute of Medical Robotics, Shanghai Jiao Tong University, Shanghai, China}
\address[2]{Shanghai Chest Hospital, Shanghai, China}
\address[3]{Imperial College London, London, UK}


\begin{abstract}
Recent evolution in deep learning has proven its value for CT-based lung nodule classification. Most current techniques are intrinsically black-box systems, suffering from two generalizability issues in clinical practice. First, benign-malignant discrimination is often assessed by human observers without pathologic diagnoses at the nodule level. We termed these data as “unsure data”. Second, a classifier does not necessarily acquire reliable nodule features for stable learning and robust prediction with patch-level labels during learning. In this study, we construct a sure dataset with pathologically-confirmed labels and propose a collaborative learning framework to facilitate sure nodule classification by integrating unsure data knowledge through nodule segmentation and malignancy score regression. A loss function is designed to learn reliable features by introducing interpretability constraints regulated with nodule segmentation maps. Furthermore, based on model inference results that reflect the understanding from both machine and experts, we explore a new nodule analysis method for similar historical nodule retrieval and interpretable diagnosis. Detailed experimental results demonstrate that our approach is beneficial for achieving improved performance coupled with faithful model reasoning for lung cancer prediction. Extensive cross-evaluation results further illustrates the effect of unsure data for deep-learning based methods in lung nodule classification.

\end{abstract}


\end{frontmatter}


\section{Introduction}
\label{sec1}

\begin{figure*}[!hbt]
\centering
\includegraphics[width = 140 mm]{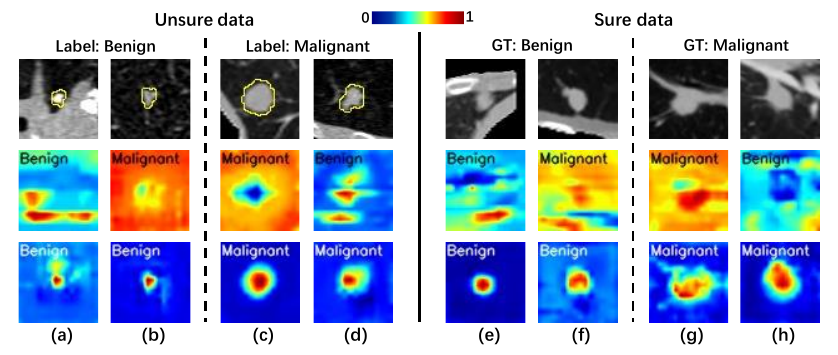}
\caption{Some examples of model interpretation. The first row shows different nodule inputs from unsure data and sure data, where the yellow contours on unsure data inputs are nodule segmentation by radiologists. The second row uses the CAM (Colormap Jet) to interpret which parts of the input contribute to the final prediction during ResNet reasoning. The third row shows the CAMs of our proposed model. Both the input images and CAMs are taken from the middle cross-sections of their 3D patches. The “Label” of unsure data is manually assigned by a malignancy score threshold 3 here. “GT” represents the ground truth of sure data that is confirmed by pathologic examination. For cases (a) and (e), although ResNet predicts “benign” on benign nodules correctly, this correct prediction comes from the misinterpreted evidence apart from the nodule regions. For cases (b) and (f), ResNet misclassifies benign nodules to “malignant”, where the CAM highlights both nodule and background regions. Our model can activate reliable features that are faithful to the nodule regions.
}
\label{fig1} 
\end{figure*}

Lung cancer is one of the major causes of cancer-related death worldwide in the last 10 years \citep{siegel2021cancer, sung2021global}. 
Screening for lung cancer with low-dose helical computed tomography (CT) has been shown in the National Lung Screening Trial (NLST) to reduce mortality from lung cancer by 20\% in high-risk individuals relative to screening with chest radiography \citep{national2011reduced}. 
Driven by CT data, deep learning is advantageous for fast computer-aided diagnosis (CAD) of lung cancer which involves lung nodule detection and benign-malignant classification. 
In this paper, rather than nodule detection, we mainly focus on nodule malignancy classification, which presents a challenge due to the diversified shapes, textures and contextual environments of lung nodules \citep{qin2021relationship}.

In recent years, many works have directed towards nodule classification based on Convolutional Neural Networks (CNNs). 
To improve the performance of nodule heterogeneity discrimination of CNN models, a diversity of approaches have been proposed, such as 
the model ensemble with multi-level inputs \citep{shen2015multi, shen2017multi, xu2020mscs,xie2017transferable, xie2018knowledge}, multi-task learning (MTL) with auxiliary tasks (e.g., nodule segmentation \citep{wu2018joint, yang2019probabilistic}, reconstruction \citep{xie2019semi}, attribute regression \citep{liu2019multi}), and relational learning from multiple nodules within a patient \citep{liao2019evaluate, yang2020relational, liu2021net} (see details in \autoref{sec::related_work_1}).
However, these studies suffer from labeling accuracy and explainability in model reasoning.




First, most of the existing methods focuses on improving nodule malignancy classification accuracy within typical publicly available dataset such as LIDC-IDRI (Lung Image Database Consortium and Image Database Resource Initiative) \citep{armato2011lung}.
During the annotation of LIDC-IDRI, the characteristics of nodules were assessed by multiple radiologists, for which the score of “likelihood of malignancy” was subjectively rated on a five-point scale
under the assumption that the CT scan was originated from a 60-year-old male smoker \citep{armato2011lung, mcnitt2007lung}, introducing inherent ambiguity problems (e.g., inter-rater variability, nodule malignancy uncertainty).
Although many attempts have been made to overcome these problems (\autoref{sec::related_work_2}), models learned from LIDC-IDRI dataset could only mimic the radiologists' reasoning by statistical fitting rather than conduct benign-malignant nodule classification in real clinical practice, since LIDC-IDRI dataset lacks definite pathologically-proven ground truth.
Other work in \autoref{sec::related_work_2} applied public data from 
TIANCHI challenges\footnote{\url{https://tianchi.aliyun.com/competition/entrance/231601/introduction/}}, NLST trial\footnote{\url{https://cdas.cancer.gov/datasets/nlst/}} \citep{national2011national, national2011reduced}, and Kaggle’s 2017 Data Science Bowl (DSB) competition (NLST subset)\footnote{\url{https://www.kaggle.com/c/data-science-bowl-2017/}} to predict lung cancer by semi-supervised learning \citep{xie2019semi} or patient-level multiple instance learning (MIL) \citep{ardila2019end,liao2019evaluate,ozdemir20193d,liu2021net}. Unfortunately, end-to-end training for nodule-level prediction was hindered by these datasets due to lack of complete annotations such as position coordinates and pathologic diagnosis of each nodule.


In this paper, we term these datasets as “unsure data” by its nature of uncertainty in subjective annotation and specific target estimation, which has been largely unaddressed by the medical image analysis community. The “unsure data” was first raised in \cite{wu2019learning} referring to the LIDC-IDRI nodules with intermediate malignancy scores, while we expand the scope of “unsure data” for all the samples without pathologically-confirmed labels at the nodule-level. This definition may have some overlaps with “noisy data (label)” and “weak label”, but “unsure data” puts more emphasis on the negative effect of the other two data definitions. To circumvent these issues, we constructed a new, high-quality annotated “sure dataset” which is described in \autoref{subsec::dataset}.

Second, in addition to the problem of unsure data, further validation and explanation are desirable to ensure the systems is trustworthy \citep{jacobs2019google}. The purpose of an explainable AI (XAI) system is to make its behavior more interpretable and credible to humans by providing explanations and evidence \citep{gunning2019xai}. Class Activation Mapping (CAM) \citep{zhou2016learning} could help to retrospectively interpret the CNN's reasoning process by performing a weighted sum of the final feature maps,
which enables a model to disclose salient information and lend insights into failure cases.

To illustrate failure cases visually, we use CAM in \autoref{fig1} to analyze some validation nodules misclassified or misinterpreted by a ResNet \citep{he2016deep} trained with unsure LIDC-IDRI dataset or our sure dataset using binary cross-entropy loss.
For instance, in the misinterpreted cases (a) and (e), the class “benign” has a bias toward background regions with high correlation as a distractor for malignancy prediction, indicating that the model may not acquire the incentive to learn benign evidence on nodule regions using training data that is only annotated with patch-level classes. However, generalization performance may degrade if the testing data does not have the same correlation \citep{zhang2021understanding}.


Evidently yet easily overlooked, failure cases in \autoref{fig1} deviate from the requirement of the nodule malignancy classification task that the reliance of the model on reliable and faithful features must be guaranteed \citep{samek2019explainable, gu2020net}, especially when the training data is small or in the cases of out-of-distribution (OoD).


Motivated by the above challenges, this paper proposes a faithful learning framework for CT-based lung cancer diagnosis based on the incorporation of both sure and unsure data. 

First of all, we design a three-branch synergic model that not only learns to classify the sure data nodules in the primary task, but also learns to diagnose like a radiologist in two auxiliary tasks for conducting nodule segmentation and malignancy score regression using unsure data. This integration approach can alleviate the negative impact of unsure data risk while elaborately adapting the unsure data knowledge for sure data learning.


Then, to endow the model with the ability to learn reliable features that are focused on the nodule regions, we leverage the CAM not only an afterthought but also a first-class citizen during training. 
To be specific, we propose a novel loss function for model online regularization called adaptive CAM-SEM-Loss (ad-CSL) by introducing “interpretability constraints” \citep{gunning2019xai} , which drives the model to express the malignancy features from the nodule regions and suppress the features in background regions under the supervision of nodule segmentation maps (SEM). 
 
In addition, based on the intra-class variance of different nodules' benign-malignant predictions (machine reasoning) and malignancy score regression outputs (mimic expert reasoning), we explore a new nodule diagnosis strategy that automatically retrieves the most similar nodules identified in a historical database, relative to a testing nodule. 
This can provide more clues and evidence for radiologists and clinicians by referring to the prior knowledge of similar historical nodule cases.



Our main contributions of this work can be summarized as follows:

\begin{enumerate}
\item A new issue for jointly learning with unsure data and our newly constructed sure data is highlighted and our work represents the first attempt for addressing this issue systematically.
\item A synergic model is proposed to integrate the unsure data knowledge with two auxiliary tasks and ultimately promote the performance of sure data classification.
\item A novel regularization scheme is proposed, which feeds back the online generated CAM to modify the classification process in such a way that the model could be more robust by learning the faithful nodule features.
\item An effective nodule diagnosis strategy is developed, which is practical for clinical usage and extensive experiments are performed to investigate the effect of unsure data and the associated problems.
\end{enumerate}





\section{Related Work}
\label{sec::related_work}

\subsection{CNN-based lung cancer prediction}
\label{sec::related_work_1}
Lung cancer prediction typically refers to the classification of benign-malignant nodules, which is indispensable to the powerful Convolutional Neural Networks (CNNs) in recent years. 
Many attempts have been done to improve the classification performance of CNN-based methods.

First, ensemble model learning is commonly used to extract the multi-level input features. \cite{shen2015multi}  proposed a weight-shared network (MCNN) to learn discriminative features from 3D nodule patches with different scales, which was then simplified by applying a multi-crop pooling strategy in MC-CNN \citep{shen2017multi}. 
Using multi-scale input, MSCS-DeepLN \citep{xu2020mscs} combined three independent sub-networks 
to generate the final ensemble prediction.
For better model nodule heterogeneity, \cite{xie2017transferable} developed a transferable ensemble model 
using three input patches characterizing overall appearance, nodule shapes, and voxel values, respectively. 
As suggested by \cite{setio2016pulmonary}, MV-KBC \citep{xie2018knowledge} extended the former work  \citep{xie2017transferable} by decomposing a 3D nodule volume onto nine fixed view planes and fed 2D patches into 27 sub-networks. 

Second, multi-task learning (MTL) helps exploit sharable knowledge in nodule-involved tasks.
Based on the LIDC-IDRI dataset, 
\cite{hussein2017risk} implicitly explored the potential of 
nodule attributes to improve the malignancy prediction by using MTL. Moreover, \cite{chen2016automatic} modeled the internal relationship between the nodule attributes and malignancy. 
Furthermore, \cite{liu2019multi} designed an MTL model that renders a mutual influence between the nodule classification and attribute score regression tasks.
Meanwhile,  \cite{wu2018joint} and \cite{yang2019probabilistic} conducted joint learning for nodule segmentation and malignancy prediction within an U-Net \citep{ronneberger2015u} structured model.


Third, relational learning may explore the incremental value of nodule data.
In the Kaggle’s 2017 Data Science Bowl (DSB) competition, the first-place model \citep{liao2019evaluate} formulated the cancer prediction as a multiple instance learning (MIL) problem that evaluated the cancer probability of a patient with multiple detected nodules. \cite{yang2020relational}  learned the relations between multiple solitary nodules from a single patient within the LIDC-IDRI dataset. Considering that some contextual features could be malignancy-related in the sense of statistics, \cite{liu2021net}  fused the features from the nodule and its surrounding structures to learn the malignancy patterns via an attention mechanism, which was also evaluated on the 2017 DSB dataset using MIL.


\subsection{Learning from unsure nodule data}
\label{sec::related_work_2}
In lung cancer prediction, most of the current work are based on one static dataset LIDC-IDRI \citep{armato2011lung}. 
Learning from this unsure dataset has several obstacles. 

First, assessed by multiple radiologists, the malignancy rating scores for each nodule may encounter a stochastic inter-observer variability. 
\cite{carrazza2016investigating}  has discovered the latent negative effect of aggregating raters’ disagreements. \cite{liao2021learning} proposed a ‘divide-and-rule’ model (MV-DAR) to learn from ambiguous labels by alleviating the value of inconsistent and unreliable nodule annotation. Nevertheless, label consensus matters in the classification task.


Second, current work commonly formulated the LIDC-IDRI nodule malignancy prediction as a binary classification task. The binary labels are crudely assigned using a simple method \citep{han2013texture} that 
based on the average malignancy score by enforcing a hard predefined threshold of score.
A large number of nodules with an average score 3 were often discarded as uncertain class.
To make use of these uncertain nodules, \cite{wu2019learning, lei2020meta, lei2021meta} applied ordinal regression to learn the relationship among the three classes. Furthermore, \cite{liu2019multi} employed a Siamese network with a margin ranking loss to model the malignancy score difference. 

Third, due to the lack of nodule-level pathologic diagnoses, models can only learn from experts' knowledge which is subjective and maybe inaccurate relative to ground truth labels. LIDC-IDRI contains a small set of cases (157 patients) with diagnosis data at the patient-level \citep{mcnitt2007lung} where four ratings (0: unknown, 1: non-malignant disease, 2: primary lung cancer, 3: metastatic lesion) were recorded along with five diagnosis methods. Based on this dataset, \cite{shen2016learning} developed a MIL framework for patient-level lung cancer prediction, indicating the potential of using definite pathologically-proven CT data for lung cancer diagnosis.


In addition to LIDC-IDRI, datasets from the ANODE09 \citep{van2010comparing} , LUNA16 \citep{setio2017validation} (a subset of LIDC-IDRI) and TIANCHI challenges are less directly used for lung cancer prediction but for nodule detection as they only provided the nodule locations and diameters. However, researchers can leverage these malignancy-unlabeled data to support the supervised classification model by semi-supervised learning \citep{xie2019semi}.
Besides, the dataset of the National Lung Cancer Screening Trial (NLST) \citep{national2011national, national2011reduced} preserved numerous CT scans in a randomized controlled trial of screening tests for lung cancer. The ground truth for cancer on the NLST dataset was biopsy or surgically confirmed, while the cancer-negative cases were only defined by a minimum of one-year follow-up. The nodule locations were not collected by the NLST trial, unfortunately. Thus, lung cancer identification using the NLST dataset \citep{ardila2019end} or Kaggle’s 2017 DSB dataset \citep{liao2019evaluate, ozdemir20193d, liu2021net} (NLST subset) rely on accurate nodule detection and segmentation in advance, which is more appropriate to be treated as a long-term patient-level prediction task.

\subsection{Interpretable nodule diagnosis}
AI systems for medical imaging diagnosis are usually applied in a black-box manner, which is highly desirable to gain the trust of users \citep{gunning2019xai, samek2019explainable}. CAM \citep{zhou2016learning}  is often used to interpret what a model has learned. The CAM visualization for nodule classifiers is not guaranteed to activate nodule regions, implying a less reliable feature learning.
These classifiers are prone to overfitting as they have the strong ability of data fitting to obtain the mapping from the patch-level input to its label, especially with limited supervision in a small dataset. The performance of OoD generalization may deteriorate when the model captures the spurious and irrelevant features within the training data domain whereas the testing data does not have the same correlation. 
Many regularization solutions have been designed to solve this issue, such as Dropout \citep{srivastava2014dropout} , Stochastic Depth \citep{huang2016deep}, Label smoothing \citep{szegedy2016rethinking}, Cutout \citep{devries2017improved}, Mixup \citep{zhang2017mixup} and Cutmix \citep{yun2019cutmix} . An alternative solution is to design different attention modules to obtain more distinguishing feature representations \citep{zhang2019attention, gu2020net}. However, these methods rely on relatively large datasets and may still locate the most discriminative region of the model’s interest in the training domain. \cite{ccc}  instead proposed a CAM-loss to constrain the embedded feature maps by minimizing the difference between class activation map and class-agnostic activation map in natural images. Inspired by this attempt, we further model the CAM map explicitly as part of the training by guiding the CAM attention to the region of human interest.


\begin{figure*}[!hbt]
\centering
\includegraphics[width = 180 mm]{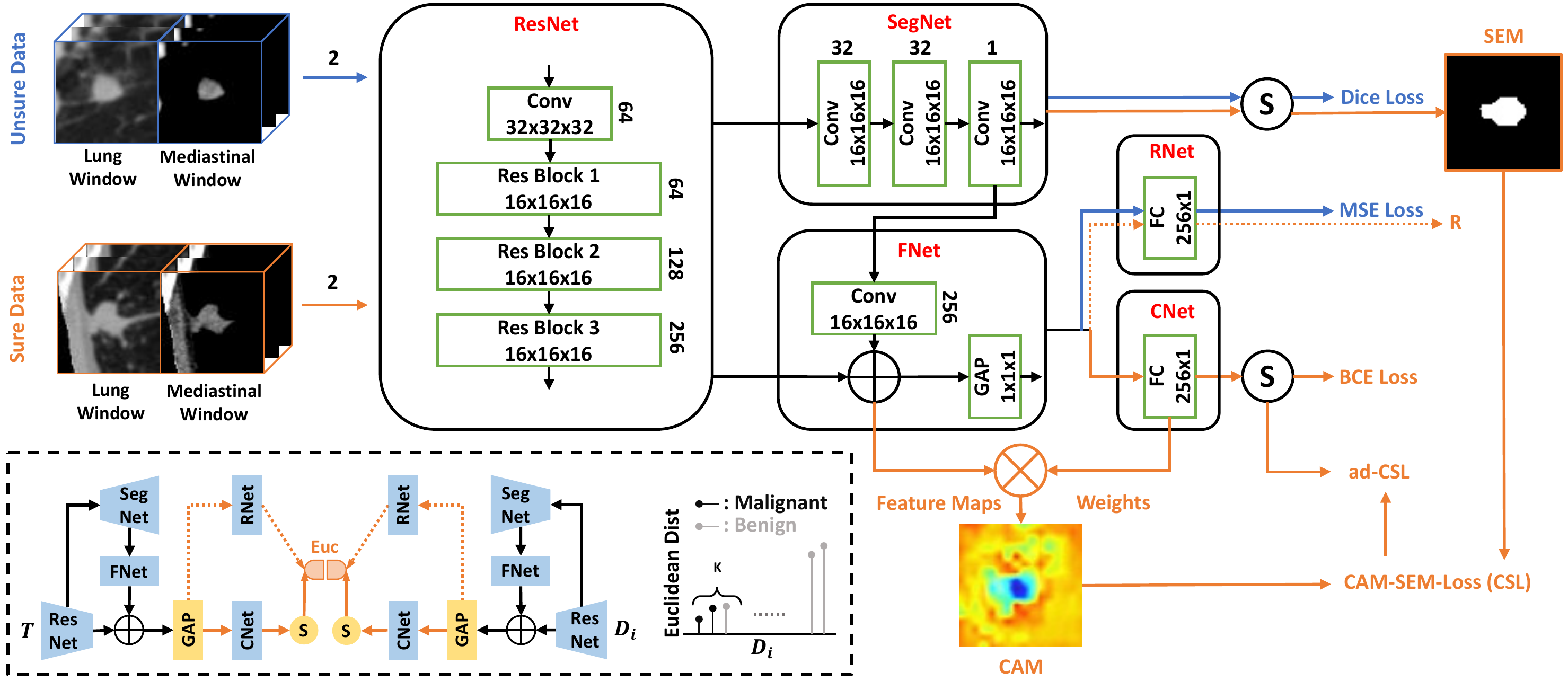}
\caption{Overview of our proposed framework for lung nodule diagnosis. 
a) The synergic model includes five modules for joint learning from unsure and sure data. Except for the last convolution layer of ResNet, SegNet and FNet that is followed by a group normalization, other convolution layers are all followed by a group normalization and a ReLU activation. ReLU activation is also performed before SegNet and after the feature addition in FNet. The resolution and channel number of each convolution or block output are denoted.
b) Adaptive CAM-SEM-Loss (ad-CSL) leverages the class activation map (CAM), nodule segmentation map (SEM) and malignant probability to regulate the synergic model for stable learning from reliable features of sure data.
c) The new nodule diagnosis method is illustrated at the bottom left corner. 
Two weight-shared pre-trained synergic models are employed to calculate the Euclidean Distance between the outputs of a nodule pair include the testing nodule $T$ and an identified nodule  $D_i$ in the historical database. The top K historical nodules with the closest Euclidean Distance relative to the testing one are selected for new malignancy assessment and diagnosis interpretation.
}
\label{fig2} 
\end{figure*}


\section{Methods}





As shown in \autoref{fig2}, the principal part of our proposed framework is a synergic model (\autoref{subsec::synergic_network}) that consists of five modules for joint learning with sure and unsure data in three tasks. In addition to the common supervisions for these three tasks, we introduce a novel regularization loss (\autoref{subsec::CAM_SEM_Loss}) for faithful nodule feature learning. 
Moreover, a new nodule diagnosis method is described in \autoref{subsec::nodule_dignosis}.
In \autoref{fig2}, the blue lines indicate operations independently performed for the stream of unsure data. The orange lines represent the data stream of sure data, where the dashed lines denote the data stream only for sure data inference in our new nodule diagnosis process. The black lines are joint canals for both data streams.

\subsection{Synergic model}
\label{subsec::synergic_network}

To better integrate the knowledge from LIDC-IDRI (unsure data) and sure data while diminishing the negative effect of domain shift in data and label space, we design a synergic model consisting of five modules (\autoref{fig2}): a backbone for feature extraction (ResNet), a branch for segmentation task (SegNet), a link bridge for feature fusion (FNet), and another two symmetric branches for nodule malignancy score regression task (RNet) and benign-malignant classification task (CNet).
The primary sure data task of classification and the auxiliary unsure data tasks of segmentation and regression are online trained simultaneously in an end-to-end manner.
The input of the synergic model is a couple of 3D patches, which are randomly selected from unsure and sure data, respectively.
The five modules of the synergic model are delved as follows:


\noindent\textbf{ResNet module:} The ResNet module for nodule feature extraction is composed of a convolutional head (size 7$\times$7$\times$7, stride 2, padding 3) and three stacked Residual blocks (two convolutional layers in each block) with the projection shortcut done by 1$\times$1$\times$1 convolution to match dimensions as used in \cite{he2016deep} .
As noticed in \cite{zhou2016learning} , the final convolutional layer before global average pooling (GAP) \citep{lin2013network}  should have a higher spatial resolution for better CAM expression. To this end, we make the following modifications as compared to the original ResNet model: 1) we remove the max-pooling layer before the first Residual block; 2) in Residual blocks 2 and 3, we set stride 1 for each convolutional operation to avoid feature down-sampling, and use dilated convolution \citep{yu2015multi}  with spacing 2 in the second Residual block and space 4 in the third Residual block to obtain the large receptive field; (3) group normalization \citep{wu2018group}  is used after each convolutional operation due to small batch size setting in the whole synergic model. Thus, the resolution of ResNet output is 16$\times$16$\times$16, which provides a reasonable representation in terms of semantic information and spatial contexture.


\noindent\textbf{SegNet module:} Our SegNet module takes inspirations from U-Net  \citep{ronneberger2015u}, a common structure utilized in most biomedical image segmentation works. 
Differently, instead of performing segmentation as the main purpose, the auxiliary SegNet module should ultimately serve the primary task of sure data nodule classification. Therefore, we give the top priority to the high demand for fast and simple module integration.
Without loss of generality, we adopt the encoder-decoder structure as the nodule segmentation implementation, where the ResNet module is regarded as a ready-made encoder to extract high-dimensional features, and the SegNet module is a lightweight decoder that utilizes the encoded features to recover the nodule segmentation distribution.
This lightweight decoder is composed of two consecutive convolutional layers with a kernel size of (3, 3, 3) for feature channel reduction (see details in \autoref{fig2}) and one convolutional layer with a (1, 1, 1) kernel to generate the distribution map of 1 channel, which is then used to output the segmentation results through a Sigmoid function. 
We do not use skip connections between ResNet and SegNet due to its light-constructed design.
Since there is no up-sampling operation in the SegNet, the spatial resolution of feature maps in this module remains unchanged after ResNet. During SegNet optimization, we down-sample the original LIDC-IDRI segmentation ground truth to the output image size by a fast zero-order spline interpolation. 
Dice coefficient loss is chosen for segmentation supervision of unsure data nodules, which is defined as

\begin{equation}
L_{seg}^{unsure}\:=\:1-\frac{2\:\sum_{i}^{N}\:y_{s,i}\:g_{s,i}\:+\:\varepsilon}{\sum_{i}^{N}\:y_{s,i}\:+\:\sum_{i}^{N}\:g_{s,i}\:+\:\varepsilon}
\end{equation}
where  $y_{s,i}$ and $g_{s,i}$ denote the predicted probability and ground truth class of the $i^{th}$ voxel belonging to the nodule region, respectively. $N$ is the number of voxels. $\varepsilon$ is a smooth factor to avoid dividing zero. Although the malignancy scores of LIDC-IDRI is unsure labels, its segmentation ground truth is relatively accurate.


\noindent\textbf{FNet module:} The FNet module integrates double-way knowledge from two sources. One stems from the output of the left ResNet module in \autoref{fig2}, which stands for high-level semantic information. The other one is generated from the upper SegNet module, introducing the distribution feature of nodule segmentation. Since the output layer of SegNet is the feature map of one channel, we apply one convolutional layer with 256 kernels (kernel size: 3$\times$3$\times$3, stride: 1, padding: 1) after the SegNet output to match the number of the ResNet output channels. With the same feature resolution and the number of feature channels, we fuse the feature maps from ResNet and segmentation-specific representations from the SegNet together by addition function, followed by a ReLU operation. 
Different from other work \citep{wu2018joint, liu2019multi}  that concatenates the feature neutrons in the fully connected structure, we directly fuse and activate the feature representation maps in the spatial dimension, which provides a straightforward interpretation mode when inferring CAM.
After performing the global average pooling (GAP), the produced features are fed into RNet and CNet for final predictions.
Such joint learning design has two benefits. On one hand, by encoding the extracted nodule segmentation information into the subsequent layers, the model localization ability and prediction performance could be enhanced. On the other hand, the gradients from RNet and CNet can also propagate to the SegNet, which can adaptively adjust the coordination between the segmentation task and prediction tasks 
while ultimately better serving the classification task.


\noindent\textbf{RNet module:} In RNet module, we formulate the nodule malignancy learning task using unsure data as a regression problem.
Compared with the commonly-used methods that simply cast this task as a binary or multiple  classification problem which assumes independence between classes, the regression approach has several benefits, such as 1) evading the unreasonable step of nodule label assignment in most of the related work in \autoref{sec::related_work}; 2) avoiding the huge waste of indeterminate data, especially in a data-hungry situation; 3) and flexibly leveraging the ordinal relationship of certain features from different malignancy scores of unsure data. 
Even though the malignancy scores may not match the true benign-malignant label, the increased suspicion level to nodule malignancy reflects the experts’ recognition in terms of the severity of nodule disease. Thus, it is of great value if the model exploits experts’ knowledge in RNet.
Also given the evidence that there is a high correspondence between the nodule malignancy score and other characteristics \citep{hussein2017risk, chen2016automatic, liu2019multi}, encoding the intrinsic malignancy relationship into model could implicitly help obtain the nodule’s heterogeneity information in size, shape and texture.
In RNet, we use a fully connected (FC) layer that outputs one neuron
to generate the unsure data malignancy score prediction. Meanwhile, RNet can also produce the experts' assessment for sure data nodules, which is vital evidence for our new nodule diagnosis strategy in \autoref{subsec::nodule_dignosis}.
We use mean square error (MSE) loss to minimize the distance error between the output value and ground truth of unsure data, which is defined as 

\begin{equation}
L_{reg}^{unsure}\:=\:\left \| y_r\:-\:g_r\right \|_{2}^{2}\:,
\end{equation}
where $y_r$ is the output of RNet and $g_r$ is the normalized malignancy score of unsure data nodule.


\noindent\textbf{CNet module:} In the primary task module of CNet, there is a FC layer that outputs one neuron followed by a Sigmoid function to generate the probability scores for nodule benign-malignant classification. The rich knowledge gained before CNet includes 1) high-level semantic information of nodule input extracted by ResNet; 2) spatial structured-features obtained from SegNet and FNet such as nodule shape, size and location; 3) and implicitly encoded features from RNet that reflect experts’ recognition to the nodule malignancy and its correlated attributes.
In this task, we use the binary cross-entropy (BCE) loss to optimize the sure data error, which is defined as

\begin{equation}
L_{cls}^{sure}\:=\:g_c\:log\:x_c\:+\:(1\:-\:g_c)\:log\:(1\:-\:x_c)\:,
\end{equation}
where $x_c$ is the output of Sigmoid function and $g_c$ is the benign-malignant label of sure data.

To resolve the large domain shift not only in data space but also in label space between unsure and sure data, we leverage a “divide-and-conquer” approach to bypass the domain conflicts after the GAP layer, so that working as an auxiliary task, the regression process will not have negative interference on the major classification task. Instead, in such a synergic way, CNet could learn more informative knowledge that single sure data supervision cannot provide.

\subsection{Learning with CAM-SEM-Loss}
\label{subsec::CAM_SEM_Loss}

Although using contextual knowledge of nodules may improve the prediction performance by data fitting algorithms, subtle statistical correlations among nodule input variables can be problematic, making models potentially more prone to overfitting on the training data.
We hypothesize that in a small medical image dataset, the object background could be regarded as a potential confounder that makes the model run against the notion of stable learning \citep{kuang2018stable}.
Thus, rather than purely fitting the observed training data, we expect to develop a predictive model that is not only robust to the wild environment changes but also faithful to the nodule regions.

In classification tasks that are supervised by patch-level annotation, CAM is usually used as a visualization tool when a model finishes optimization, but few studies fed it back to the training progress \citep{ccc}.
To alleviate the underperformance of model faithfulness in visual interpretability, in \autoref{fig2}, we propose to construct a new loss function for model regularization, called CAM-SEM-Loss (CSL), by leveraging the online generated CAM with the supervision of nodule segmentation maps (SEM). The detailed description is shown below.


For a given nodule input in each training batch, the FNet finally uses ReLU to activate the fused feature maps. 
Let $f_{k}\:(x,y,z)$ represents the activation of unit $k$ at 3D spatial location $(x,y,z)$. Then, for unit $k$ with length $L$, width $W$ and height $H$, the result of performing GAP, $F_k$, is $\frac{1}{L\:\times\: W \:\times\:H}\:{\sum}_{x,y,z}\:f_{k}\:(x,y,z)$. Thus, the input to the Sigmoid function after CNet, $S$, is ${\sum}_{k}\:{\omega}_{k}\:F_{k}$, where ${\omega}_{k}$ is the weight of CNet's fully connected layer corresponding to the unit $k$ before backpropagation and optimization of each batch.
Essentially, ${\omega}_{k}$ indicates the current positive relevance of  $F_k$ for the predicted malignant class or the negative relevance for the benign class.
Finally, the output of the Sigmoid function, $P$, is given by $\frac{1}{1\:+\:e^{(-S)}}$ to represent the nodule malignancy score.
By plugging $F_k$ into the malignancy score $S$, we obtain

\begin{equation}
\begin{aligned}
     S\: & = \:\frac{1}{L\:\times\: W \:\times\:H}\:\sum_{k}\:\omega_k\:\sum_{x,y,z}\:f_{k}\:(x,y,z)\:\\
     & = \:\frac{1}{L\:\times\: W \:\times\:H}\:\sum_{x,y,z}\:\sum_{k}\:\omega_k\:f_{k}\:(x,y,z)\:\\
\end{aligned}
\end{equation}

We define $CAM$ as the class activation map by forward propagation, where each spatial element is given by

\begin{equation}
     CAM\:(x,y,z)\: = \: \sum_{k}\:\omega_k\:f_{k}\:(x,y,z)\:\\
\end{equation}

Thus, $S\:= \:\frac{1}{L\:\times\: W \:\times\:H}\:\sum_{x,y,z}\:CAM\:(x,y,z)$, where $CAM\:(x,y,z)$ directly indicates the malignancy attention at spatial location $(x,y,z)$.
Since we use Sigmoid as our final activation function, in the attention area of class activation map, $CAM\:(x,y,z)$ will have a lower value if the model outputs a low malignancy score and a higher value for a high malignancy score.

To standardize the CAM visualization, we introduce the malignancy score $P$ into the original $CAM$. The new $CAM_c$ for predicted class $c$ is defined by

\begin{equation} \label{equation}
     CAM_c\:(x,y,z)\: = (P\:-\: Threshold)\: \sum_{k}\:\omega_k\:f_{k}\:(x,y,z)\:,
\end{equation}
where $Threshold$ is the division point to classify the benign-malignant nodule, which is normally set to 0.5 and under the following relation

\begin{equation}
c\:=\:
\left\{\begin{aligned}
 &0\:(benign)& P  <Threshold\\ 
 &1\:(malignant)& P \geqslant Threshold
\end{aligned}\right.
\end{equation}

The min-max normalization is finally used to rescale each element of $CAM_c$ with $CAM_c\:(x,y,z)\in \left [0,1 \right ]$.

Besides, in the nodule segmentation map, we define $SEM\:(x,y,z)\in \left (0,1 \right )$ as the foreground-background prediction of the pixel at the location $(x,y,z)$. By performing Sigmoid on SegNet outputs, most $SEM\:(x,y,z)$ would quickly distribute near either 0 or 1 after the training starts, providing the high confident guidance to the updating $CAM_c\:(x,y,z)$ with the information of whether this position belongs to nodule regions or not.
This inspires us to leverage the $SEM$ knowledge and give a interpretability constraint to $CAM_c$ by regulating more attention on nodule regions during training.
For this purpose, we first obtain the approximate average CAM values of nodule regions and background regions, which are defined by

\begin{equation}
AvgCAM_{ndl}=\frac{\sum_{x,y,z}\:CAM_c\:(x,y,z)\:SEM\:(x,y,z)}{\sum_{x,y,z}\:SEM\:(x,y,z)}
\end{equation}

\begin{equation}
AvgCAM_{bkg}=\frac{\sum_{x,y,z}\:CAM_c\:(x,y,z)\:(1\:-\:SEM\:(x,y,z))}{\sum_{x,y,z}(1\:-\:SEM\:(x,y,z))}
\end{equation}
where $AvgCAM_{ndl}\in \left [0,1 \right ]$, $AvgCAM_{bkg}\in \left [0,1 \right ]$.

Then, to drive the model to learn more discriminative feature representations from a nodule perspective, we formulate the CAM-SEM-Loss as follows
\begin{equation}
L_{CSL}^{sure}\:=\:max\left \{ \:0,\:AvgCAM_{bkg}-AvgCAM_{ndl} + \delta \:  \right \},
\end{equation}
which enforces $AvgCAM_{ndl}\: \geqslant \: AvgCAM_{bkg}+\delta $ in the sure data training progress, where $\delta$ is the margin parameter that adjusts the attention bias between nodule and background regions. Moreover, $L_{CSL}^{sure}$ does not neglect the contextual information of a nodule because feature learning from the background regions is not inhibited.

Considering that the CAM-SEM-Loss would treat equally to those predictions with different malignant probability scores, we further extend CAM-SEM-Loss with the knowledge of uncertainty that the prediction with a lower confidence score ($P$ around $Threshold$) should gain less supervision from CAM-SEM-Loss. To this end, the adaptive CAM-SEM-Loss (ad-CSL) is designed by multiplying the $l_1$ distance between $Threshold$ and $P$ over CAM-SEM-Loss, which is given by:

\begin{equation}
     L_{ad-CSL}^{sure} = 2\:\left \| P \:- \:Threshold \right \|_{\:l_1} \: L_{CSL}^{sure}\:,
\end{equation}
where the coefficient is 2 to match the value range of CAM-SEM-Loss.

In summary, the total loss for training includes four losses in our framework:

\begin{equation}
  L_{total} = \: L_{cls}^{sure} \: + \: \alpha \: L_{ad-CSL}^{sure} \: + \: \beta \: L_{seg}^{unsure} \: + \: \gamma \:  L_{reg}^{unsure} \:,
\end{equation}
where $\alpha$, $\beta$ and $\gamma$ are three hyper-parameters to balance these terms which are all set to 1 in our study.





\subsection{Nodule diagnosis with synergic model}
\label{subsec::nodule_dignosis}
 
We establish a new nodule diagnosis method by a retrieval algorithm based on the prior nodule information.
Specifically, using the trained model J, we first obtain the outputs for a testing nodule and each prior identified nodule $i$ in the historical database, defined as $T$ and $D_i$, respectively. Generally, the historical sure data are also used for training the synergic model if nodule locations are available.

From the perspective of synergic model, the outputs of each nodule have three forms including 1) the nodule classification score from CNet, represented as the machine inference; 2) the nodule regression score from RNet, originated from the expert knowledge; and 3) the concatenation of 1) and 2).

Then, we rank the Euclidean Distance of outputs between $T$ and each $D_i$, and acquire the ranking list of top $K$ nodules with the closest Euclidean Distance to the testing nodule, which is defined by

\begin{equation}
List_{i}^{K} = Rank^{K} \left \{ \left | T,\:D_{i} \right |_{Euc},i=1,2,...,num_D \right \},
\end{equation}
where $num_{D}$ is the number of reference nodules in the historical database.

Afterward, a new diagnosis score of the testing nodule is awarded by averaging the labels of top $K$ reference cases in $List_{i}^{K}$, which is given by 

\begin{equation}
Diag=\frac{1}{K}\sum_{j=i}Label_{j}, i\in List_{i}^{K},
\end{equation}
where $ Diag\in \left [0,1 \right ]$ and K is empirically set to 20 in this study.

By reading and matching these closely related historical nodule cases which lead to similar outcomes, clinicians can acquire more evidence and clues for the testing nodule diagnosis.

\section{Experiments}


\subsection{Datasets}
\label{subsec::dataset}


\noindent\textbf{Unsure data:}
The unsure dataset used in this study comes from the LIDC-IDRI database \citep{armato2011lung}, which consists of 1018 CT scans with annotations of the nodules given by multiple radiologists. According to the practice in LUNA16 \citep{setio2017validation} , CT scans with a slice thickness smaller than 3 mm were included (888 CTs), where nodules $\ge$ 3 mm and $<$ 30 mm accepted by at least 3 radiologists were considered as positive samples (1186 nodules). On top of that, we only involve the 919 majority solid nodules (average texture score = 5).
To alleviate the inter-observer variability as regards to malignancy voting, we first discard the nodules with the mean absolute difference (MAD) \citep{yitzhaki2003gini}  among malignancy scores larger than 0.6 and calculate the average malignancy scores for the remaining 686 nodules. 
In total, the number of nodules with an average score of 1, 2, 3, 4 and 5 is 88, 101, 338, 100 and 59, respectively. Different from any other work, further label assignment is not required in our method.

To obtain the nodule segmentation label, we first fill the internal area of the nodule boundary delineated by each radiologist. Then a 50\% consensus criterion \citep{kubota2011segmentation} is adopted to generate a single ground-truth boundary. That is, if the current voxel is annotated by two or more radiologists, the voxel point will be regarded as a nodule label. Otherwise, it will be labeled as background.


\noindent\textbf{Sure data:}
The sure dataset consists of 330 solid nodules (165 benign/165 malignant) collected from 317 patients’ CT scans retrospectively in Shanghai Chest Hospital. 
The collection and analysis of image data were approved by the ethical committee of the Shanghai Chest Hospital and adhered to the tenets of the Declaration of Helsinki.
The age of patients ranges from 25 to 81 years with an average of 57.38 ($\pm$ 11.45) years. CT scans in this dataset were acquired by multiple manufacturers of GE Medical Systems, Philips and United Imaging Health (UIH), where the slice thickness ranges from 0.625 to 3.0 mm with an average of 1.14 ($\pm$ 0.32) mm and the pixel spacing varied from 0.34 to 0.98 mm with an average of 0.58 ($\pm$ 0.22) mm.
All the cases were annotated to a single definite class (benign or malignant) diagnosed by pathological-proven examination via surgical resection. The last eligible CT scans before surgery were chosen to enable a small time gap between nodule images and ground truth. 
The exact spatial coordinate and radius of each nodule were annotated by two board-certified radiologists with more than five years of clinical experience and confirmed by one senior radiologist with more than fifteen years of clinical experience.
In this study, we only include the nodules with diameter less than 30 mm (consistent with the accepted upper limit of nodule size \citep{hansell2008fleischner}) and larger than or equals to 3 mm (lower limit for practical consideration \citep{armato2004lung}). 

Note that although the public LUNGx Challenge dataset \citep{kirby2016lungx} also provided some sure data, we do not include this dataset in our study due to its small quantity and nonuniform assessment for benign nodules. 
\begin{table*}[t]
\centering
\caption{Quantitative performance of the synergic model for the ultimate sure data nodule benign-malignant classification task by 5-fold cross-validation (under the threshold of 0.5), including combination with different modules, comparison with attention mechanism and different loss functions.}
\label{tab:table1}
\footnotesize 
\resizebox{\textwidth}{!}{%
\begin{tabular}{l|ll|l|l|lllllll|l}
\hline
\multirow{2}{*}{} & \multicolumn{2}{l|}{Tasks} & \multirow{2}{*}{Modules} & \multirow{2}{*}{\begin{tabular}[c]{@{}l@{}}Params\\ ($\times10^{6}$)\end{tabular}} & \multicolumn{7}{l|}{Results ($\%$) (mean $\pm$ standard deviation)} & \multirow{2}{*}{vis$^\star$} \\ \cline{2-3} \cline{6-12}
 & \multicolumn{1}{l|}{Sure} & Unsure &  &  & \multicolumn{1}{l|}{Sensitivity} & \multicolumn{1}{l|}{Specificity} & \multicolumn{1}{l|}{Precision} & \multicolumn{1}{l|}{Precision$_b$} & \multicolumn{1}{l|}{Accuracy} & \multicolumn{1}{l|}{AUC} & F1-score &  \\ \hline
A & \multicolumn{1}{l|}{Cls} & - & C & 3.6282 & \multicolumn{1}{l|}{64.24  $\pm$  4.85} & \multicolumn{1}{l|}{58.79 $\pm$   8.70} & \multicolumn{1}{l|}{61.28  $\pm$  4.36} & \multicolumn{1}{l|}{62.05  $\pm$  3.92} & \multicolumn{1}{l|}{61.52  $\pm$  4.02} & \multicolumn{1}{l|}{69.53 $\pm$   3.39} & 62.54  $\pm$  3.18 & $\checkmark$ \\ \hline
B & \multicolumn{1}{l|}{Cls} & Seg & C,S & 3.8770 & \multicolumn{1}{l|}{67.27 $\pm$   9.27} & \multicolumn{1}{l|}{64.85 $\pm$   14.16} & \multicolumn{1}{l|}{66.93  $\pm$  9.89} & \multicolumn{1}{l|}{66.49  $\pm$  7.32} & \multicolumn{1}{l|}{66.06  $\pm$  7.46} & \multicolumn{1}{l|}{74.67  $\pm$  8.39} & 66.47 $\pm$   6.74 & - \\ \hline
C & \multicolumn{1}{l|}{Cls} & Reg & C,R & 3.6284 & \multicolumn{1}{l|}{66.67 $\pm$   6.64} & \multicolumn{1}{l|}{67.27  $\pm$  10.03} & \multicolumn{1}{l|}{67.72  $\pm$  6.02} & \multicolumn{1}{l|}{67.04   $\pm$ 3.55} & \multicolumn{1}{l|}{66.97  $\pm$  4.00} & \multicolumn{1}{l|}{76.80  $\pm$  2.90} & 66.84  $\pm$  3.68 & - \\ \hline
D & \multicolumn{1}{l|}{Cls} & Seg+Reg & C,S,R & 3.8773 & \multicolumn{1}{l|}{68.48  $\pm$  8.48} & \multicolumn{1}{l|}{\textbf{69.70}  $\pm$  3.83} & \multicolumn{1}{l|}{\textbf{69.29}  $\pm$  1.98} & \multicolumn{1}{l|}{69.35  $\pm$  4.91} & \multicolumn{1}{l|}{69.09  $\pm$  3.26} & \multicolumn{1}{l|}{76.51   $\pm$ 2.96} & 68.64 $\pm$   4.94 & $\checkmark$ \\ \hline
E & \multicolumn{1}{l|}{Cls} & Seg+Reg & C,S,R,ARL$^{*}$ & 3.8773 & \multicolumn{1}{l|}{69.70  $\pm$  6.06} & \multicolumn{1}{l|}{67.27  $\pm$  8.44} & \multicolumn{1}{l|}{68.57  $\pm$  4.55} & \multicolumn{1}{l|}{69.15  $\pm$  2.80} & \multicolumn{1}{l|}{68.48  $\pm$  2.61} & \multicolumn{1}{l|}{76.95 $\pm$   6.43} & 68.80  $\pm$  2.44 & $\checkmark$ \\ \hline
F & \multicolumn{1}{l|}{Cls} & Seg+Reg & C,S,R,F & 3.8842 & \multicolumn{1}{l|}{75.15  $\pm$  4.02} & \multicolumn{1}{l|}{63.03 $\pm$   11.08} & \multicolumn{1}{l|}{67.69 $\pm$   5.04} & \multicolumn{1}{l|}{71.63  $\pm$  1.15} & \multicolumn{1}{l|}{69.09  $\pm$  3.66} & \multicolumn{1}{l|}{77.23   $\pm$ 6.23} & 70.94 $\pm$   1.43 & $\checkmark$ \\ \hline
G & \multicolumn{1}{l|}{Cls} & Seg+Reg & C,S,R,F,ARL$^{*}$ & 3.8842 & \multicolumn{1}{l|}{71.52  $\pm$  3.09} & \multicolumn{1}{l|}{66.06  $\pm$  9.07} & \multicolumn{1}{l|}{68.28  $\pm$  6.76} & \multicolumn{1}{l|}{69.62  $\pm$  4.45} & \multicolumn{1}{l|}{68.79  $\pm$  5.47} & \multicolumn{1}{l|}{76.38  $\pm$  5.03} & 69.75 $\pm$   4.46 & - \\ \hline
H & \multicolumn{1}{l|}{Cls (CL$^{\#}$)} & Seg+Reg & C,S,R,F & 3.8842 & \multicolumn{1}{l|}{72.73  $\pm$  4.29} & \multicolumn{1}{l|}{62.42  $\pm$  6.24} & \multicolumn{1}{l|}{66.09  $\pm$  4.06} & \multicolumn{1}{l|}{69.59  $\pm$  3.97} & \multicolumn{1}{l|}{67.58  $\pm$  3.90} & \multicolumn{1}{l|}{77.34  $\pm$  5.14} & 69.17  $\pm$  3.51 & $\checkmark$ \\ \hline
I & \multicolumn{1}{l|}{Cls (CSL)} & Seg+Reg & C,S,R,F & 3.8842 & \multicolumn{1}{l|}{71.52   $\pm$ 7.81} & \multicolumn{1}{l|}{64.85  $\pm$  12.36} & \multicolumn{1}{l|}{68.16  $\pm$  6.83} & \multicolumn{1}{l|}{69.80  $\pm$  2.97} & \multicolumn{1}{l|}{68.18  $\pm$  3.46} & \multicolumn{1}{l|}{77.37   $\pm$ 5.91} & 69.15  $\pm$  2.67 & $\checkmark$ \\ \hline
J & \multicolumn{1}{l|}{Cls (ad-CSL)} & Seg+Reg & C,S,R,F & 3.8842 & \multicolumn{1}{l|}{\textbf{76.97} $\pm$   4.11} & \multicolumn{1}{l|}{64.24  $\pm$  9.27} & \multicolumn{1}{l|}{68.67  $\pm$  5.67} & \multicolumn{1}{l|}{\textbf{73.42}  $\pm$  4.87} & \multicolumn{1}{l|}{\textbf{70.61} $\pm$ 5.30} & \multicolumn{1}{l|}{\textbf{77.65}  $\pm$  5.64} & \textbf{72.46}  $\pm$  4.20 &  $\checkmark$ \\ \hline
\multicolumn{12}{@{}l}{$^{\star}$ CAMs shown in \autoref{fig3}  \qquad $^{*}$ ARL--Attention Residual Learning \citep{zhang2019attention}\qquad  $^{\#}$ CL--CAM-Loss \citep{ccc} } 
\end{tabular}
}
\end{table*}

\subsection{Data preprocessing}
Considering that the valid nodules are restricted inside the two lungs, we think it reasonable to first perform robust lung segmentation of chest CT scans as a prerequisite for automated nodule analysis, which is conducted by: 1) binarization using OTSU \citep{otsu1979threshold} as threshold selection method; 2) extraction of the largest connected components A; 3) hole filling on A to obtain B; 4) coarse lung segmentation C by subtracting A from B; 5) denoising on C to obtain D; 6) closing operation on D using non-flat morphological structuring element to obtain E; 7) binarization on E using manually set threshold (0.5) to obtain the robust segmented lung mask F; 8) multiplication on the raw CTs and the mask F.
Steps 6) and 7) are key guarantees for robust lung segmentation where abnormalities are often present. Otherwise, many juxta-pleural nodules will be missed due to erroneous lung segmentation \citep{armato2004automated}.

We clipped the Hounsfield unit (HU) values into the lung window interval [-1000, 400 HU] and mediastinal window interval [-160, 240 HU], respectively. Then each window is normalized to the linear range of [0, 1]. To resolve the anisotropic nature of CTs, the 3D images were resampled to a fixed 0.5 mm/voxel along all three axes using spline interpolation. 
To extract 3D patches, we first consider a cube of $64\times64\times64$ voxel, which could completely enclose the nodule with a diameter less than 30mm.
Then we concatenate the cubes from the lung and mediastinal window to obtain the 2-channel nodule input.

The nodule volume augmentation method includes random flipping over the three axes, random rotation around the three axes with angles chosen from 90°, 180°, or 270°, and random transposing by reversing or permuting between every two axes of the 3D image.

\subsection{Experiment setting}
All the experiments are implemented in PyTorch \citep{paszke2019pytorch}  with a single NVIDIA GeForce GTX 1080 Ti GPU and learned using Adam optimizer \citep{kingma2014adam}  with the initialized learning rate of 1e-3 and the maximum epoch number of 100 (batchsize = 1). The iteration times in each epoch is based on the number of sure training data.
The validation set occupies 20\% of the training set in each experiment to monitor the performance of training model.
All the experiments and results involving or having involved the training of sure data are strictly conducted by 5-fold cross-validation, in which four subsets were employed to train the model, one subset was used as testing data.

\subsection{Evaluation metrics}
To evaluate the model performance comprehensively, our evaluation metrics include Sensitivity (Recall), Specificity (also called Recall$_b$, if treating benign as positive sample), Precision, Precision$_b$ (Precision in benign class), Accuracy with the cut-off value of 0.5, AUC (area under the receiver operating characteristic curve) and F1-score.

\subsection{Quantitative evaluation of multi-task learning in synergic model}

In this subsection, we conduct ablation study and comparison experiments for quantitative evaluation of sure data performance with synergic model. 
To observe the contributions of different modules in our synergic model, we first combine different modules based on model A.
As the numerical results illustrated in \autoref{tab:table1}, model A, consisting of a ResNet (backbone) and a CNet, can hardly obtain the malignancy discrimination ability with only the sure data knowledge. This is probably restricted by the limited amount of sure data that cannot drive the model to extract the malignancy-related features supervised by definite nodule labels.
Model B, C and D first attempt to integrate the sure and unsure data using an online co-training network in nodule malignancy classification task. 
Although there is a large domain shift between two datasets (in both data and label space), additional learning with unsure data nodules simultaneously in auxiliary tasks, model B, C and D outperform model A in all the evaluation metrics by an average 4.7\%, 5.6\% and 7.3\%, respectively. Such significant promotion indicates the potential of leveraging unsure data knowledge for sure data discrimination. Among these three structures, model D obtains the superimposing positive effects of adding SegNet and RNet modules.
Based on model D, model F bridges the nodule segmentation features to the end of backbone for both unsure and sure data, achieving better performance in Sensitivity, Precision$_b$, AUC and F1-score.

\begin{figure}[!hbt]
\centering
\includegraphics[width = 80 mm]{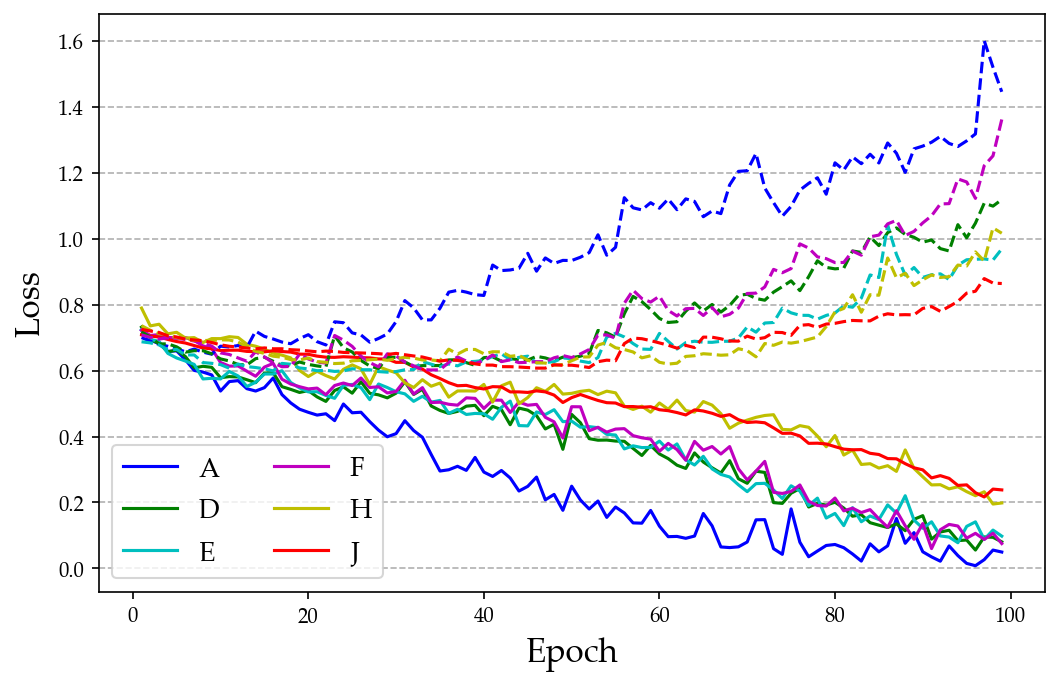}
\caption{Binary cross-entropy (BCE) loss error vs. epoch for different models in \autoref{tab:table1}) in training (80\%) and validation (20\%) performances. Solid lines represent training losses and dashed lines denote validation losses.
}
\label{fig9} 
\end{figure}

\begin{figure*}[!hbt]
\centering
\includegraphics[width = 183 mm]{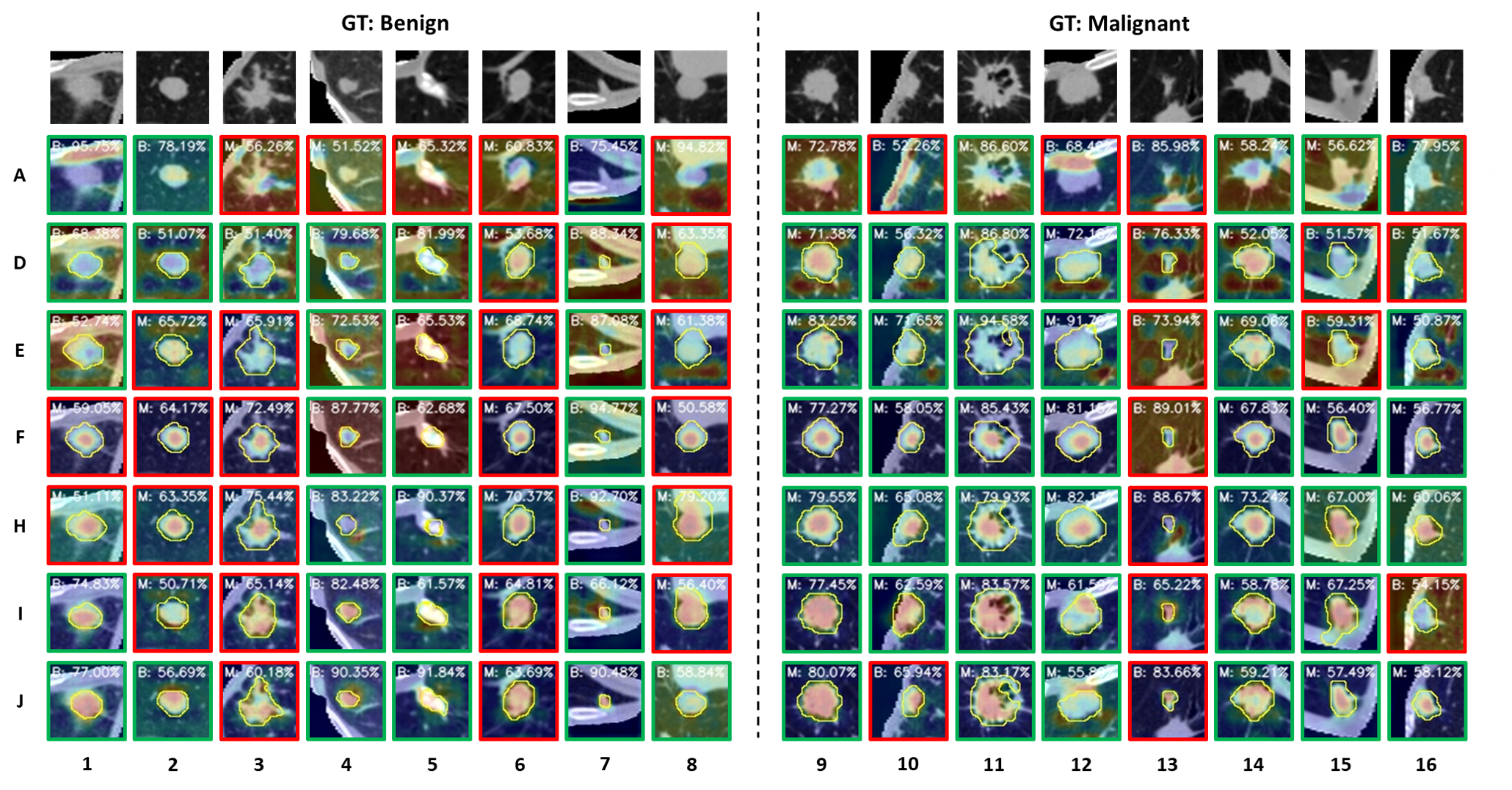}
\caption{Visualization of CAMs obtained by models chosen from \autoref{tab:table1} which are A, D (A + SegNet + RNet), E (D + ARL), F (D + FNet), H (F + CL), I (F + CSL) and J (F + ad-CSL). The first row shows the nodule inputs from sure data. "B" and "M" are the benign and malignant prediction results under the threshold of 0.5. The scores are the probabilities for each predicted class.
The yellow contours on each image depict the automatic segmentation of sure data nodules generated from SegNet followed by Sigmoid function (threshold: 0.5). 
Green or red border to an image denotes the true or mistaken prediction, respectively. Both the input images and CAMs are taken from the central slices of their 3D patches.
}
\label{fig3} 
\end{figure*}

We choose two typical existing techniques that are possible to facilitate nodule heterogeneity discrimination.
The first technique is the attention mechanism. We replace the Residual blocks of model D and F by ARL (attention residual learning) blocks which work well for 2D skin lesion classification \citep{zhang2019attention}. As suggested by \cite{zhang2019attention}, we apply spatial attention which is more suitable than channel attention and mixed attention to help focus on semantic regions of a target object.
As the results of model E and G shown in \autoref{tab:table1}, spatial ARL block has limited improved performance or local degradation compared with model D and F.
One possible reason is that attention mechanisms like \cite{chen2017sca, hu2018squeeze} would further enhance the correlated feature learning inside the model which may not fit for small-sample tasks.
The second technique is CAM-loss (CL) \citep{ccc} which regulates the feature maps by minimizing the difference between CAM and class-agnostic activation map (CAAM) in natural image classification tasks. We test the effect of CL based on model F. The results of model H show that except AUC, other metrics drop comprehensively when applied this extra loss in our task. We explain this phenomenon in the next \autoref{subsec::Qualitative}.

To evaluate our loss function, we first use CAM-SEM-Loss (CSL) in model I. Compared to model F, model I appears a performance degradation which is similar to model H. Under such circumstances, an upturn happened to model J with adaptive CAM-SEM-Loss (ad-CSL) by encoding the uncertainty of nodule prediction to CSL, which outperforms any other model broadly, especially for Sensitivity, Precision$_b$, Accuracy and F1-score. In addition, \autoref{fig9} shows that, model J has a higher training error but lower validation error in BCE loss for sure data classification compared with other models, which indicates that our method has a positive effect on reducing overfitting.



Up to the present, these quantitative results in \autoref{tab:table1} affirm the superiority of aggregating multi-task modules and ad-CSL loss for discriminating high-correlated features. Most published papers ended the experiments here, which constrained the analytical insight instead of advancing it. As a matter of fact, 
quantitative evaluation might be inadvertently involved in a hidden fraud that the nodule malignancy prediction is not derived from the reasonable evidence for model reasoning, but rather from the outcomes of data fitting.

\subsection{Qualitative evaluation of synergic model}
\label{subsec::Qualitative}

For further visualization analysis, we adopt the class activation mapping (CAM) for sure data classification to reveal the attention regions identified by different models in \autoref{tab:table1}.
Because all the models are constructed in a common structure ending with a GAP and a fully connected layer 
, we apply \autoref{equation} followed by a min-max normalization to express the interpretation results in \autoref{fig3}.

As illustrated in model A, there is no guarantee for a common black-box CNN model to learn 
nodule-relevant features with limited knowledge and supervision. As a result, the small amount of sure training data causes overfitting quickly since an early stage, as illustrated in \autoref{fig9}.
Based on the cases of A-1, A-10 and A-12 in  \autoref{fig3}, we can interpret that model A may remember the juxta-pleural feature of benign nodules during training but it suffers from domain bias when the model is applied on testing data.

With the incorporation of unsure data knowledge, model D appears some regular patterns of CAMs that are guided by the nodule segmentation map. As marked by the yellow contour lines in \autoref{fig3}, sure data nodules can be well automatically segmented by the light-weight SegNet branch which is trained by unsure data. Although there is no direct shortcut that introduces the segmentation knowledge to the end of backbone in model D, an effective constraint can be exhibited in D-1, D-10 and D-12.
However, in some CAMs, model D not only highlights the features in nodule regions but also activates massive background regions because nodule segmentation learns both foreground and background information.
Nevertheless, jointly learned with sure and unsure data, model D obtained a significant improvement in \autoref{tab:table1} but implicitly failed in interpretation performance.

Compared to model D, we observe a worse CAM result in model E that the spatial ARL block \citep{zhang2019attention} aggravates the incorrect concentration on background for benign predictions (e.g., E-1, E-4, E-5, E-13 and E-15) and reduces the nodule feature learning for malignant prediction (e.g., E-6, E-8 and E-9).
This indicates that attention techniques may compound the preconception error if the correct features are not guaranteed in advance.

\begin{figure*}[!hbt]
\centering
\includegraphics[width = 180 mm]{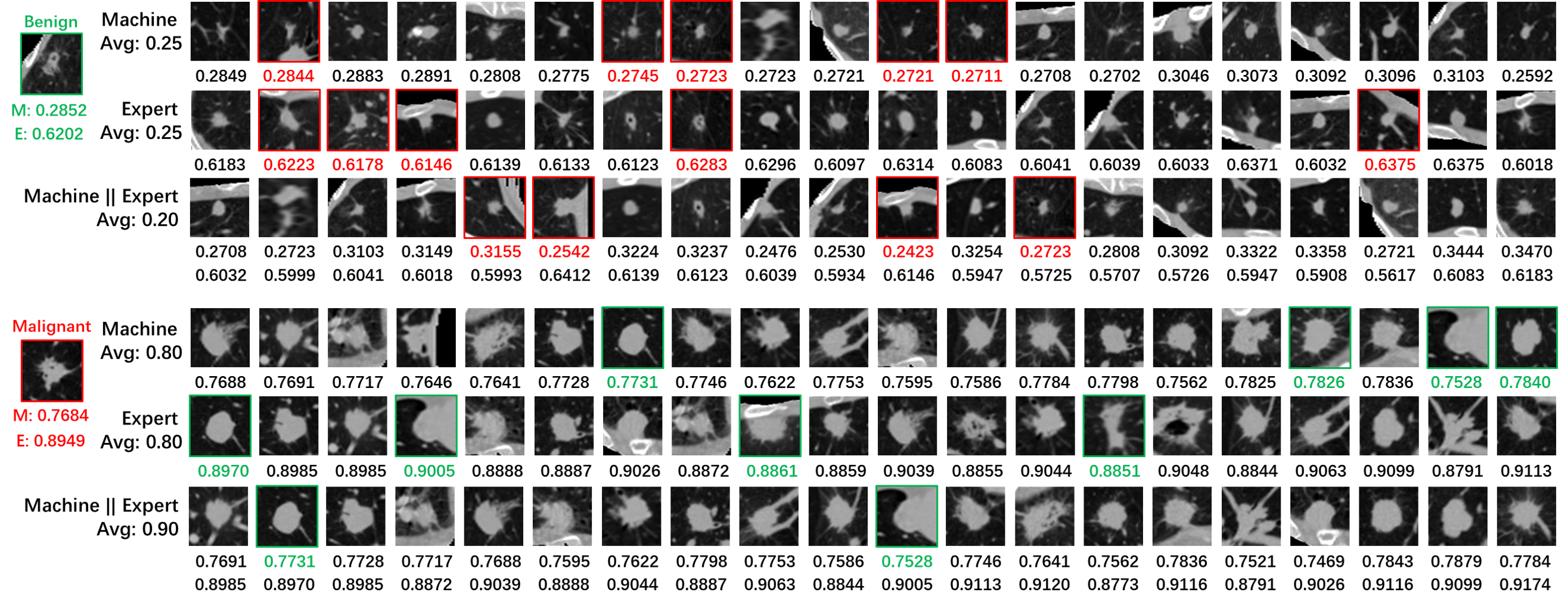}
\caption{Illustration of similar nodule retrieval (K=20) using the new computer-aided diagnosis (CAD) method. The upper half shows similar nodules relative to a testing nodule whose ground truth is benign while the lower half tests on a malignant nodule. For machine reasoning, the scores below each image denote the malignant probability generated by CNet followed by a Sigmoid. For expert reasoning, the scores represent the normalized malignancy score regression output from RNet. Red or green border to an image denotes the wrong retrieved class for benign or malignant nodule, respectively. 
}
\label{fig4} 
\end{figure*}

For model F that bridges the SegNet features to the final convolution layer of model D using an FNet, its visual saliency maps preserve an evident involvement of nodule segmentation knowledge, whereas most benign predictions preserve similar unexpected attention as the model D and E.
Added with CAM-loss \citep{ccc} which was applied in natural image classification tasks, model H could erase part of the background attention of benign predictions, but it failed to highlight their nodule regions. This may be caused by the inherent differences between classification tasks using natural images and nodule images. In contrast to natural images that possess large training samples with obvious class-discriminative features, 1) nodule malignancy classification is a one-object and two-category task; 2) the benign-malignant information of sure nodule data is visually interchangeable even for experimented radiologists; 3) as a carrier for radiomics, the representation of a CT scan is restricted by its image acquisition mode that invasive pathological knowledge may not be captured; 4) sure nodule data-hungriness issue makes CNNs unreliable; 5) the attention maps cannot serve as reliable priors for CAM-Loss in nodule task (CAM-loss is under the assumption that their attention maps can still serve as reliable priors for tasks).

By incorporating CSL loss, a powerful interpretability constraint designed for nodule images, the model I can put more emphasis on the feature variables in nodule area either for malignant predictions or benign predictions (e.g., I-4, I-7 and I-13).
Benefiting from the encoding of prediction uncertainty, the model J optimized with additional ad-CSL presents a stronger attention ability that its produced CAM maps are highly calibrated with nodules, and therefore make more reliable predictions in \autoref{tab:table1} and faithful feature representations in \autoref{fig3}. This is attributed to the key role of adaptive strategy in ad-CSL that enables the model to first strengthen discrimination ability and segmentation performance if a nodule prediction is of low confidence while focusing more on semantic information of nodules.
Note that, it does not matter if the prediction is incorrect because the wrong predictions should also have the correct CAMs.


\subsection{Evaluation of new nodule diagnosis strategy}

We hypothesize that the variance information of intra-class predictions can bring about the similarity of a nodule pair in different feature spaces and these predictions are able to represent the overall assessment of a nodule characteristic such as malignant probability in CNet and malignancy score in RNet. Based on this hypothesis, \autoref{subsec::nodule_dignosis} proposed a new computer-aided diagnosis (CAD) strategy that retrieves the most similar nodules in a historical database and generates another diagnosis score for testing nodules.

\begin{table}[]
\caption{Performance of new nodule diagnosis strategy by treating the new diagnosis score as malignant probability and conducting evaluation under the same metrics on sure data (\%).}
\label{tab:table2}
\footnotesize 
\setlength{\tabcolsep}{1.5mm}{
\begin{tabular}{clllllll}
\hline\hline
Reasoning & \multicolumn{1}{c}{Sen} & \multicolumn{1}{c}{Spe} & \multicolumn{1}{c}{Pre} & \multicolumn{1}{c}{Pre$_b$} & \multicolumn{1}{c}{Acc} & \multicolumn{1}{c}{AUC} & \multicolumn{1}{c}{F1} \\ \hline\hline
Machine (CNet) & 76.97 & 62.42 & 68.22 & 72.48 & 69.7 & 76.22 & 72.01 \\
Expert (RNet) & 76.97 & 64.85 & 69.17 & 73.38 & 70.91 & 77.07 & 72.73 \\
Machine $\|$ Expert & \textbf{78.18} & \textbf{65.45} & \textbf{70.65} & \textbf{74.15} & \textbf{71.82} & \textbf{77.67} & \textbf{73.88} \\ \hline\hline
\end{tabular}
}
\end{table}

\begin{figure*}[!hbt]
\centering
\includegraphics[width = 183 mm]{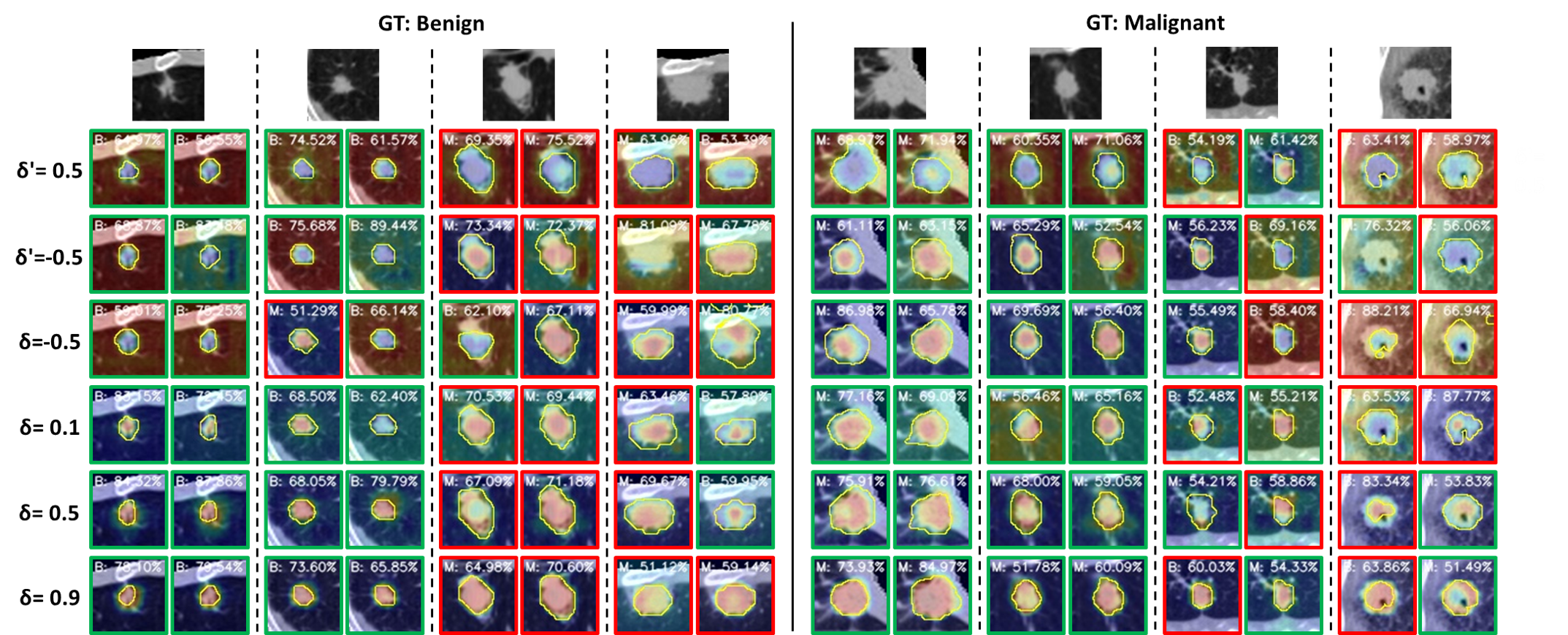}
\caption{Comparison of CAMs generated by models optimized with CSL or ad-CSL using different margin parameters. $\delta$ denotes the margin parameter with a bias toward nodule regions while $\delta^{'}$ represents that with a bias toward background regions. CAMs in the left and right columns of each nodule image come from the model using CSL and ad-CSL, respectively.
}
\label{fig5} 
\end{figure*}

Under such a diagnosis strategy, we treat the new diagnosis score as malignant probability and evaluate using the same metrics for testing nodules. As shown in \autoref{tab:table2}, based on model J, the performance of machine reasoning is slightly worse than that of expert reasoning, which is broadly in line with the performance of model J in \autoref{tab:table1} using the traditional diagnosis method.
This indicates the potential to leverage the expert knowledge from unsure domain in testing phase although we do not align the distribution between sure data and unsure data in RNet.
After adding the conditions for the judgment of similarity by concatenating the cognition from machine and experts, one can achieve better performance of nodule classification, even surpassing its original model J.
Furthermore, prior knowledge of historical nodule cases can be reused as an assistant for nodule diagnosis in clinical practice. We present an intuitive nodule diagnosis results in \autoref{fig4}. Other than the binary prediction that most studies only considered, we additionally provide the top 20 similar nodules as diagnosis reference that making the CAD system more user-friendly and practical.


\section{Discussion}

\subsection{Comparing different modes for RNet}

We compare the influence of different modes for RNet in \autoref{tab:table3} based on model C and model D that both apply the RNet to learn malignancy scores of unsure data.
First, cross-entropy (CE) loss is commonly used for multi-class discrimination tasks. However, this loss function has never been applied to the five-category classification task of LIDC-IDRI malignancy scores, which is often formulated as binary classification.
The second mode (Ord) applies ordinal regression for RNet to model the ranking of malignancy scores. For this mode, we use the effective method from \cite{diaz2019soft} that converts unsure data labels into soft probability distributions pairing with cross-entropy loss. For these two modes, we modify the output neuron number of the FC layer to 5 and add a Softmax function to generate the prediction in RNet.

\begin{table}[h]
\caption{Comparison between models with different modes for RNet (\%).}
\label{tab:table3}
\footnotesize 
\setlength{\tabcolsep}{1.5mm}{%
\begin{tabular}{cclllllll}
\hline \hline
Model &
  Reg/Cls &
  \multicolumn{1}{c}{Sen} &
  \multicolumn{1}{c}{Spe} &
  \multicolumn{1}{c}{Pre} &
  \multicolumn{1}{c}{Pre$_b$} &
  \multicolumn{1}{c}{Acc} &
  \multicolumn{1}{c}{AUC} &
  \multicolumn{1}{c}{F1} \\ \hline \hline
\multirow{3}{*}{C} & CE  & 68.48 & 63.03 & 65.37 & 67.65 & 65.76 & 72.95 & 66.33 \\
                        & Ord & 68.48 & 62.42 & 64.51 & 66.63 & 65.45 & 72.36 & 66.39 \\
                        & MSE & 66.67 & 67.27 & 67.72 & 67.04 & 66.97 & 76.80  & 66.84 \\ \hline
\multirow{3}{*}{D}   & CE  & 67.88 & 67.27 & 67.94 & 67.94 & 67.58 & 77.02 & 67.52 \\
                        & Ord & 67.88 & 66.67 & 67.15 & 67.48 & 67.27 & 75.02 & 67.48 \\
                        & MSE & 68.48 & 69.70  & 69.29 & 69.35 & 69.09 & 76.51 & 68.64 \\ \hline \hline
\end{tabular}
}
\end{table}

The result in \autoref{tab:table3} shows that the regression mode (MSE) obtained the best performances in both model C and model D. The traditional cross-entropy loss, which fails to model the ordinal relationship among five malignancy classes, has worse overall performances compared to the regression mode (MSE), especially in model C. We also observe that ordinal regression mode (Ord) is less likely to have good cooperation with the major task during optimization, leading to the obstruction of sure data learning. Thus, regression mode is a simple and effective way to leverage the ordinal variables that resided in unsure data and finally contribute to the sure data classification. The results in \autoref{tab:table3} also demonstrate that the segmentation task can improve the model discrimination ability comprehensively under the comparison between model C and model D.

\subsection{Margin parameter $\delta$}
\label{subsec::parameter}

In this experiment, we explore the influence of margin parameter $\delta$ in CAM-SEM-Loss (CSL) and  adaptive CAM-SEM-Loss (ad-CSL). As can be seen in \autoref{tab:table4} and \autoref{fig5}, to better illustrate the working mechanism of these two loss functions, we add mode 1 ($bkg \geqslant ndl + \delta^{'}$) that enable the model to learn more discriminative features in the background, whose CSL is formulated as

\begin{equation} \label{equation15}
L_{CSL}^{sure}\:=\:max\left \{ \:0,\: AvgCAM_{ndl} - AvgCAM_{bkg} + \delta^{'} \:  \right \},
\end{equation}

Driven by \autoref{equation15}, we successfully control the CAM to focus on the background in most cases of $bkg  \geqslant  ndl + 0.5$ in \autoref{fig5}. 
Although the degradation of quantitative performance appears for ad-CSL in \autoref{tab:table4}, there is an interesting phenomenon that CSL can still maintain acceptable evaluation results.
This indicates that learning from the background can still realize data fitting to some extent when no reliable features are provided.

\begin{table}[t]
\caption{Performances of models using CAM-SEM-Loss (CSL) or ad-CSL with different margin parameters $\delta/\delta^{'}$ in two modes ($\%$).}
\label{tab:table4}
\footnotesize 
\setlength{\tabcolsep}{1mm}{
\begin{tabular}{cccccccccc}
\hline\hline
Mode & $\delta / \delta^{'}$ & Loss & Sen & Spe & Pre & Pre$_b$ & Acc & AUC & F1 \\ \hline\hline
\multirow{4}{*}{\begin{tabular}[c]{@{}c@{}}bkg\\ $\geqslant$ \\ ndl+$\delta^{'}$ \end{tabular}} & \multirow{2}{*}{0.5} & CSL & 67.27 & 68.48 & 68.38 & 67.68 & 67.88 & 77.36 & 67.68 \\
 &  & ad-CSL & 51.52 & 72.73 & 71.5 & 61.85 & 62.12 & 76.8 & 53.29 \\ \cline{2-10} 
 & \multirow{2}{*}{-0.5} & CSL & 70.91 & 61.21 & 65.48 & 68.25 & 66.06 & 76.11 & 67.26 \\
 &  & ad-CSL & 73.94 & 68.48 & 70.07 & 72.87 & 71.21 & 75.74 & 71.80 \\ \hline
\multirow{8}{*}{\begin{tabular}[c]{@{}c@{}} ndl\\ $\geqslant$ \\ bkg +$\delta$ \end{tabular}} & \multirow{2}{*}{-0.5} & CSL & 73.33 & 64.24 & 67.51 & 70.80 & 68.79 & 77.04 & 70.14 \\
 &  & ad-CSL & 73.33 & 63.64 & 67.05 & 70.37 & 68.48 & 75.45 & 69.98 \\ \cline{2-10} 
 & \multirow{2}{*}{0.1} & CSL & 71.52 & 63.64 & 66.57 & 68.94 & 67.58 & 76.58 & 68.87 \\
 &  & ad-CSL & 76.97 & 64.85 & 68.83 & 73.89 & 70.91 & 77.76 & 72.54 \\ \cline{2-10} 
 & \multirow{2}{*}{0.5} & CSL & 71.52 & 64.85 & 68.16 & 69.80 & 68.18 & 77.37 & 69.15 \\
 &  & ad-CSL & 76.97 & 64.24 & 68.67 & 73.42 & 70.61 & 77.65 & 72.46 \\ \cline{2-10} 
 & \multirow{2}{*}{0.9} & CSL & 64.85 & 73.94 & 72.05 & 67.94 & 69.39 & 77.12 & 67.79 \\
 &  & ad-CSL & 81.82 & 58.79 & 67.42 & 77.35 & 70.30 & 77.59 & 73.30 \\ \hline\hline
\end{tabular}
}
\end{table}

\begin{table*}[t]
\caption{Performances of model A, model D and model D+ad-CSL trained with inputs with different modalities (\%).}
\label{tab:table5}
\footnotesize 
\resizebox{\textwidth}{!}{%
\begin{tabular}{lllllllll}
\hline\hline
Model & Modality & Sensitivity & Specificity & Precision & Precision$_b$ & Accuracy & AUC & F1-score \\ \hline\hline
\multirow{4}{*}{Model A} & 64 & 64.24 $\pm$ 4.85 & 58.79 $\pm$ 8.70 & 61.28 $\pm$ 4.36 & 62.05 $\pm$ 3.92 & 61.52 $\pm$ 4.02 & 69.53 $\pm$ 3.39 & 62.54 $\pm$ 3.18 \\
 & x & 66.06 $\pm$ 14.01 & 61.21 $\pm$ 14.65 & 63.69 $\pm$ 7.57 & 65.45 $\pm$ 7.16 & 63.64 $\pm$ 6.57 & 72.58 $\pm$ 2.64 & 64.00 $\pm$ 8.04 \\
 & x-resize-64 & 63.64 $\pm$ 11.18 & 56.97 $\pm$ 5.88 & 59.49 $\pm$ 2.23 & 61.88 $\pm$ 4.99 & 60.30 $\pm$ 3.24 & 65.25 $\pm$ 4.24 & 61.10 $\pm$ 6.21 \\
 & x-padding-64 & 69.70 $\pm$ 10.14 & 64.24 $\pm$ 7.02 & 66.19 $\pm$ 3.23 & 68.74 $\pm$ 5.87 & 66.97 $\pm$ 3.76 & 72.40 $\pm$ 3.11 & 67.52 $\pm$ 5.40 \\ \hline
\multirow{4}{*}{Model D} & 64 & 68.48 $\pm$ 8.48 & 69.70 $\pm$ 3.83 & 69.29 $\pm$ 1.98 & 69.35 $\pm$ 4.91 & 69.09 $\pm$ 3.26 & 76.51 $\pm$ 2.96 & 68.64 $\pm$ 4.94 \\
 & x & 69.70 $\pm$ 9.58 & 61.21 $\pm$ 6.18 & 64.11 $\pm$ 5.68 & 67.36 $\pm$ 7.02 & 65.45 $\pm$ 6.17 & 73.13 $\pm$ 3.93 & 66.64 $\pm$ 7.16 \\
 & x-resize-64 & 69.09 $\pm$ 2.27 & 64.24 $\pm$ 7.52 & 66.23 $\pm$ 4.77 & 67.37 $\pm$ 2.67 & 66.67 $\pm$ 3.71 & 70.69 $\pm$ 5.21 & 67.53 $\pm$ 2.68 \\
 & x-padding-64 & 72.12 $\pm$ 9.47 & 63.03 $\pm$ 9.47 & 66.31 $\pm$ 5.73 & 69.73 $\pm$ 7.21 & 67.58 $\pm$ 6.25 & 72.97 $\pm$ 5.47 & 68.83 $\pm$ 6.47 \\ \hline
\multirow{4}{*}{\begin{tabular}[c]{@{}c@{}}Model D\\ +\\ ad-CSL\end{tabular}} & 64 & 73.33 $\pm$ 4.85 & 66.67 $\pm$ 9.58 & 69.23 $\pm$ 5.16 & 71.39 $\pm$ 3.45 & 70.00 $\pm$ 4.33 & 76.01 $\pm$ 5.35 & 71.01 $\pm$ 3.39 \\
 & x & 76.36 $\pm$ 6.47 & 57.58 $\pm$ 8.57 & 64.50 $\pm$ 5.05 & 71.01 $\pm$ 5.64 & 66.97 $\pm$ 5.20 & 72.21 $\pm$ 4.02 & 69.79 $\pm$ 4.78 \\
 & x-resize-64 & 69.70 $\pm$ 8.57 & 63.03 $\pm$ 11.08 & 66.05 $\pm$ 4.57 & 68.13 $\pm$ 4.24 & 66.36 $\pm$ 2.61 & 72.51 $\pm$ 3.88 & 67.30 $\pm$ 2.64 \\
 & x-padding-64 & 67.27 $\pm$ 9.85 & 66.06 $\pm$ 8.44 & 66.75 $\pm$ 4.15 & 67.51 $\pm$ 5.47 & 66.67 $\pm$ 3.95 & 72.47 $\pm$ 2.35 & 66.58 $\pm$ 5.23 \\ \hline\hline
\end{tabular}
}
\end{table*}

In normal circumstances, we regulate $\delta$ with positive values in mode 2 ($ndl \geqslant bkg + \delta$), where ad-CSL is prone to have greater results than CSL in overall quantitative evaluation \autoref{tab:table4}.
When $\delta$ rises to 0.9, an aggressive parameter that forces the model to pay nearly all the concentration on nodule regions, its salient visualization map will present a “cleaner” attention in the background regions compared to $\delta = 0.1$ and $\delta = 0.5$ for both CSL and ad-CSL in \autoref{fig5}. 
This action guarantees the high faithful feature learning from nodule regions, but it will cause performance fluctuation, especially for Sensitivity and Specificity.

While in cases of $ndl \geqslant bkg + 0.1$ and $ndl \geqslant bkg + 0.5$ that obtain the top overall quantitative performance in \autoref{tab:table4}, we can see in \autoref{fig5}  that CSL and ad-CSL still consider the feature learning from nodule background as we formulate (ad-) CSL based on the relativity between foreground and background instead of simply penalizing the CAM values of background to zero.

Besides, we evaluate the performance when setting $\delta/\delta^{'}$ with a negative value -0.5 for both modes. Such a conservative parameter setting may have few impacts on CAM results which show the a similar attention pattern compared with model F in \autoref{fig3}.

Overall, we give the conclusion for this experiment: 1) $\delta$ can adjust the attention weight between nodule regions and background regions; 2) ad-CSL performs better than CSL in quantitative evaluation when guiding attention to the nodule regions; 3) there is an attention balance between nodule and its background for good lung cancer prediction; 4) the quantitative performance cannot reflect the reliability of model learning, misleading the observers accidentally.

\subsection{Influence of using different input modalities}
\label{subsec::modalities}

\begin{figure}[!hbt]
\centering
\includegraphics[width =70 mm]{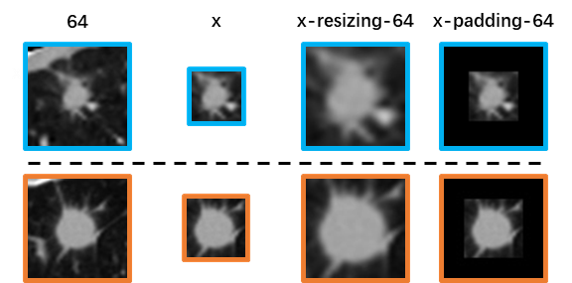}
\caption{Sample images of sure data with different input modalities. Up: benign nodule; down: malignant nodule.}
\label{fig6} 
\end{figure}

We argue that nodule inputs matter for the whole deep learning process.
As shown in \autoref{fig6}, we consider four input modalities that could be used for nodule benign-malignant classification task: 1) the original $64\times64\times64$ voxel size cube (64); 2) the nodule size cube (x) that is cropped according to the nodule coordinate and radius; 3) a new $64\times64\times64$ voxel size cube (x-resizing-64) that is generated by resizing the nodule size cube using cubic interpolation; and 4) another $64\times64\times64$ voxel size cube (x-padding-64) that is produced by performing zero padding on the nodule size cube. For unsure data, the ground truth of the segmentation map follows the transformation of its nodule data. Modalities of x and x-resizing-64 can be approximately regarded as nodule-level inputs.

We choose Model A and Model D in this experiment because FNet seems redundant for the other three new input modalities that they have already possessed the knowledge of nodule size and location. We additionally conduct ad-CSL ($\delta = 0.5$) based on model D to evaluate the effect of interpretability constrain on these different inputs.

As shown in \autoref{tab:table5}, it can be observed that training with model D leads to a broad growth relative to model A, indicating the significant value of integrating unsure data knowledge for each input. 
If adding ad-CSL during model D training, the performance of using original input 64 could also get promoted even without FNet.
However, ad-CSL can hardly empower model D with improved overall quantitative performance for input x-resizing-64 and input x-padding-64 in Accuracy and F1-score. Meanwhile, the model fed with input x produces more false-positive predictions.
It is not difficult to give the reason for such a situation. If a kind of input has already embedded the function of nodule region expression and background suppression into itself, ad-CSL would lose its value.
Therefore, learning with more nodule semantic information, model A fed with input x and input x-padding-64 can outperform the same model fed with input 64 by only using sure data.

\begin{figure*}[!hbt]
\centering
\includegraphics[width = 180 mm]{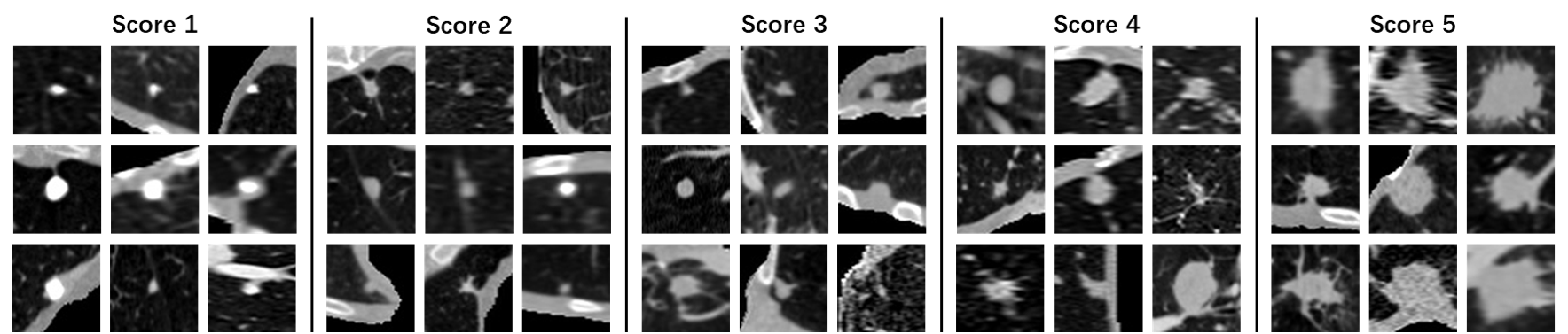}
\caption{Sample images of LIDC-IDRI (unsure data) nodules. Nodules with different average malignant scores are randomly selected after pre-processing and illustrated here.
}
\label{fig7} 
\end{figure*}

\begin{table*}[t]
\caption{Performances of cross-evaluation between sure \& unsure data and different integration methods of two datasets (\%). "ft" denotes the mode of transfer learning that pre-training model using unsure data and fine-tuning model using sure data. The column "Scenario" represents the different label assignment methods (benign/malignant) for unsure data based on their average malignant scores (five-point scale).}
\label{tab:table6}
\resizebox{\textwidth}{!}{%
\begin{tabular}{lllllllllll}
\hline\hline
Module & Scenario & \begin{tabular}[c]{@{}l@{}}Training\\ data \& mode\end{tabular} & \begin{tabular}[c]{@{}l@{}}Testing\\ data\end{tabular} & Sensitivity & Specificity & Precision & Precision$_b$ & Accuracy & AUC & F1-score \\ \hline\hline
\multirow{2}{*}{RNet} & \multirow{2}{*}{} & Unsure & Sure & 95.15 ± 3.64 & 24.85 ± 4.45 & 55.90 ± 1.32 & 85.27 ± 10.21 & 60.00 ± 2.06 & 74.77 ± 6.06 & 70.39 ± 1.53 \\
 &  & Unsure+sure & Sure & 73.94 ± 9.88 & 61.82 ± 12.21 & 66.58 ± 6.86 & 70.77 ± 7.51 & 67.88 ± 6.46 & 75.37 ± 5.75 & 69.59 ± 6.12 \\ \hline
\multirow{20}{*}{CNet} & \multirow{4}{*}{1/2345} & Unsure & Unsure & 96.49 ± 0.64 & 80.85 ± 7.37 & 97.16 ± 1.10 & 77.34 ± 2.46 & 94.47 ± 0.70 & 97.49 ± 1.22 & 96.82 ± 0.39 \\
 &  & Unsure & Sure & 100.00 ± 0.00 & 1.21 ± 2.42 & 50.31 ± 0.63 & 100.00 ± 0.00 & 50.61 ± 1.21 & 60.75 ± 8.80 & 66.94 ± 0.55 \\
 &  & Unsure+sure & Sure & 90.30 ± 8.44 & 13.94 ± 6.80 & 51.16 ± 1.95 & 64.79 ± 20.94 & 52.12 ± 3.40 & 62.26 ± 6.66 & 65.23 ± 3.44 \\
 &  & Unsure+sure(ft) & Sure & 64.85 ± 8.04 & 67.27 ± 9.66 & 66.95 ± 4.71 & 65.88 ± 4.37 & 66.06 ± 4.13 & 72.87 ± 5.56 & 65.49 ± 4.45 \\ \cline{2-11} 
 & \multirow{4}{*}{12/345} & Unsure & Unsure & 90.56 ± 5.80 & 52.92 ± 3.25 & 83.48 ± 0.80 & 71.18 ± 14.41 & 80.18 ± 3.65 & 80.97 ± 2.82 & 86.80 ± 2.78 \\
 &  & Unsure & Sure & 100.00 ± 0.00 & 0.61 ± 1.21 & 50.15 ± 0.31 & 100.00 ± 0.00 & 50.30 ± 0.61 & 67.09 ± 6.14 & 66.80 ± 0.27 \\
 &  & Unsure+sure & Sure & 84.24 ± 6.47 & 33.94 ± 13.61 & 56.47 ± 3.20 & 68.20 ± 1.97 & 59.09 ± 3.71 & 69.99 ± 2.95 & 67.33 ± 0.82 \\
 &  & Unsure+sure(ft) & Sure & 67.88 ± 6.53 & 66.67 ± 12.71 & 67.92 ± 6.85 & 67.27 ± 4.40 & 67.27 ± 5.72 & 73.02 ± 6.63 & 67.52 ± 4.67 \\ \cline{2-11} 
 & \multirow{4}{*}{12/45} & Unsure & Unsure & 76.15 ± 4.92 & 87.23 ± 6.88 & 84.19 ± 7.28 & 81.44 ± 2.78 & 82.17 ± 3.14 & 90.53 ± 1.73 & 79.65 ± 3.22 \\
 &  & Unsure & Sure & 89.09 ± 5.28 & 31.52 ± 12.80 & 56.91 ± 3.75 & 75.38 ± 6.27 & 60.30 ± 4.43 & 70.74 ± 5.40 & 69.25 ± 2.01 \\
 &  & Unsure+sure & Sure & 76.97 ± 1.48 & 57.58 ± 8.57 & 64.77 ± 4.63 & 71.05 ± 4.20 & 67.27 ± 4.66 & 73.90 ± 4.10 & 70.27 ± 3.13 \\
 &  & Unsure+sure(ft) & Sure & 73.33 ± 5.55 & 60.00 ± 7.02 & 64.82 ± 5.24 & 69.22 ± 6.06 & 66.67 ± 5.50 & 74.75 ± 6.37 & 68.77 ± 5.03 \\ \cline{2-11} 
 & \multirow{4}{*}{123/45} & Unsure & Unsure & 58.47 ± 4.78 & 95.42 ± 2.41 & 80.09 ± 8.71 & 88.40 ± 1.23 & 86.86 ± 2.41 & 87.48 ± 1.86 & 67.41 ± 5.43 \\
 &  & Unsure & Sure & 80.00 ± 5.94 & 52.73 ± 9.51 & 63.14 ± 4.78 & 72.44 ± 6.50 & 66.36 ± 5.37 & 71.31 ± 6.94 & 70.42 ± 4.30 \\
 &  & Unsure+sure & Sure & 69.09 ± 13.19 & 69.70 ± 13.28 & 70.51 ± 7.24 & 70.17 ± 6.35 & 69.39 ± 5.86 & 76.64 ± 5.27 & 68.80 ± 7.73 \\
 &  & Unsure+sure(ft) & Sure & 67.88 ± 5.94 & 66.06 ± 8.44 & 67.02 ± 5.11 & 67.35 ± 3.73 & 66.97 ± 4.22 & 73.72 ± 6.61 & 67.24 ± 4.11 \\ \cline{2-11} 
 & \multirow{4}{*}{1234/5} & Unsure & Unsure & 45.45 ± 11.76 & 97.13 ± 1.09 & 60.33 ± 9.36 & 95.01 ± 0.92 & 92.70 ± 1.26 & 93.54 ± 3.33 & 51.11 ± 10.23 \\
 &  & Unsure & Sure & 32.73 ± 14.77 & 93.94 ± 3.32 & 85.94 ± 9.12 & 58.74 ± 4.97 & 63.33 ± 6.74 & 74.97 ± 6.77 & 45.13 ± 17.09 \\
 &  & Unsure+sure & Sure & 59.39 ± 12.94 & 70.91 ± 10.43 & 67.51 ± 6.39 & 64.27 ± 6.61 & 65.15 ± 6.06 & 71.55 ± 7.25 & 62.42 ± 8.38 \\
 &  & Unsure+sure(ft) & Sure & 69.70 ± 8.99 & 61.21 ± 14.26 & 65.02 ± 8.15 & 66.70 ± 7.42 & 65.45 ± 7.81 & 71.72 ± 7.99 & 66.88 ± 6.91 \\ \hline\hline
\end{tabular}%
}
\end{table*}

In this experiment, several phenomena were also revealed: 1) the background of a nodule could also contribute to the extraction of discriminative nodule features corresponding to the final prediction (64 vs. x-padding-64); 2) only learning from the CT information in the nodule regions can hardly achieve prominent malignancy classification performance (e.g., x, x-resizing-64), indicating that there may not exist CT-based manifestation of pathologies for nodule malignancy identification.

\subsection{Cross-evaluation and other integration methods of two datasets}
\label{sec::discussion_4}

In this subsection, we first investigate the effect of adopting a malignancy classifier on unsure data through cross-evaluation of two datasets and then study on other data integration methods.
As can be seen in \autoref{tab:table6}, this study mainly uses two model structures that apply RNet module and CNet module on ResNet backbone, respectively.

For the first structure with RNet, we conduct regression task based on two datasets and optimize the model by only using MSE loss. Given the fact that the ground truth of sure data has a high confidence level, we align the benign-malignant labels of sure data to the lowest-highest malignancy scores of unsure data. The model was separately trained on the unsure data only and on a mixed data consisting of both datasets.

The second structure is the same as model A that is designed for classification task. Thus, in this task, we first identified 5 scenarios by setting different division thresholds to assign binary labels for unsure data. The samples of unsure data nodules with different average malignancy scores are shown in \autoref{fig7}.
In each scenario, we conducted four experiments: 1) model A was independently trained and tested on only unsure data with 5-fold cross-validation; 2) model A was trained on the whole unsure data and tested on sure data; 3) model A was trained on a merged two datasets. 4) model A was pre-trained using unsure data (50 epochs; learning rate:  1e-3) and fine-turned using sure data on the whole pre-trained model (50 epochs; learning rate: 1e-4).

From \autoref{tab:table6}, observing the results of experiments that conducted both training and testing using unsure data only, we can find that good evaluation performance can be achieved within the unsure data domain, especially for scenario 1/2345 and scenario 12/45 which removed the uncertain score. According to the nodules samples shown in \autoref{fig7}, most samples with average score 1 are calcified nodules, which have significant visible differences from other nodules with higher scores.
The differences between nodules with scores 1\&2 and those with scores 4\&5 are distinct as well.
As far as we know, scenario 12/45 was broadly used in many work for binary classification of LIDC-IDRI nodules. Compared between scenario 12/345 and scenario 123/45, better overall classification performance could be achieved by grouping the uncertain score into benign class, which is consistent with the experimental results in \citep{han2015texture} . This paper concluded that these uncertain nodules are more similar to benign ones. However, this inner-dataset experimental conclusion is still under suspicion as the ground truth for LIDC-IDRI is not available.


Once the scenario 12/45 is evaluated using sure data, model trained with unsure data tends to generate many false-positive predictions while most evaluation results fall sharply, especially for Specificity. Evidence reveals that monitored by sure testing data, false-positive problem could be alleviated successively when the division threshold moves from low-score side to high-score side, indicating the ordinal relationship of unsure data malignancy score in the view of sure data.

Mixed data learning and transfer learning \citep{zhang2020learning} are two effective methods for multi-data integration. As shown the results with the training mode of "Unsure + sure" that two datasets are fed into one task model simultaneously, improved performance could be achieved either for regression task or classification task.
Moreover, transfer learning that gets the model fine-tuned could further correct the domain bias.
However, these two methods will inevitably encounter the thorny problem of unsure data label assignment as well as knowledge waste of a large uncertain subset.

\begin{table}[]
\caption{Performance comparison of different synergic model structures (Scenario: 12/45)  and other state-of-the-art results for lung nodule classification using LIDC-IDRI dataset (\%) .}
\label{tab:table7}
\footnotesize 
\setlength{\tabcolsep}{0.7mm}{
\begin{tabular}{cllllllll}
\hline\hline
\multicolumn{1}{l}{} & Methods & Sen & Spe & Pre & Pre$_b$ & Acc & AUC & F1 \\ \hline\hline
\multirow{7}{*}{Others} &  \citep{shen2017multi} & 77.00 & 93.00 & - & - & 87.14 & 93.00 & - \\
 &  \citep{hussein2017risk} & - & - & - & - & 91.26 & - & - \\
 &  \citep{xie2018fusing} & 84.40 & 90.88 & 82.09 & - & 88.73 & 94.02 & 83.23 \\
 &  \citep{xie2017transferable} & 83.83 & 94.56 & 88.40 & - & 91.01 & 95.35 & 86.07 \\
 &  \citep{xie2018knowledge}& 86.52 & 94.00 & 87.75 & - & 91.60 & 95.70 & 87.13 \\
 &  \citep{xie2019semi} & 84.94 & 96.28 & - & - & 92.53 & 95.81 & - \\
 &  \citep{xu2020mscs} & 85.58 & 95.87 & 90.39 & - & \textbf{92.64} & 94.00 & 87.91 \\ \hline
\multicolumn{1}{l}{\multirow{4}{*}{Ours}} & C & 76.15 & 87.23 & 84.19 & 81.44 & 82.17 & 90.53 & 79.65 \\
\multicolumn{1}{l}{} & C,S & 79.90 & 90.97 & 88.34 & 84.38 & 85.91 & 92.99 & 83.83 \\
\multicolumn{1}{l}{} & C,S,F & \textbf{88.10} & 90.94 & 89.53 & \textbf{90.40} & 89.65 & 95.23 & 88.57 \\
\multicolumn{1}{l}{} & C,S,F,ad-CSL & 85.54 & \textbf{96.81} & \textbf{95.85} & 88.85 & 91.67 & \textbf{96.10} & \textbf{90.38} \\ \hline\hline
\end{tabular}%
}
\end{table}

\subsection{The classification performance of synergic model and comparison with state-of-the-art using unsure dataset}


\autoref{tab:table7} displays the results of our models (Scenario: 12/45) and other state-of-the-art methods using only LIDC-IDRI dataset.
The 5-fold cross-validation results show that our synergic model structure (without the application of RNet for regression task) can also be applied to unsure data for its nodule benign-malignant determination (binary classification).  
Compared within our different models,
the module SegNet and module FNet can make a significant contribution in terms of each evaluation metric. When the model with SegNet and FNet is additionally optimized by ad-CSL loss function ($\delta = 0.5$) , it can achieve much higher Specificity and Precision but lower Sensitivity and Precision$_b$, indicating that ad-CSL ($\delta = 0.5$) works more effectively for the prediction of benign nodules which are defined by unsure data. 

Compared with other state-of-the-art methods, our model obtain the best results in Specificity, Precision, AUC and F1-score on minority samples with small batch size and little parameter tuning.
Nevertheless, comparison between our model and state-of-the-art methods is beyond the scope of this study that mainly contributes to the integration of two datasets' knowledge and regularization method for faithful feature learning. More importantly, unsure data should deliver the dominant position to sure data under the consideration of scientific strictness and clinical validity.

In addition, there probably be a trade-off between good quantitative results and faithful nodule feature learning for unsure data, where the malignancy scores could be only derived from visible features by multiple radiologists whose decisions are made under the assumption that all the CT scans belong to a 60-year-old male smoker. 
We hence argue that without the pathologically-proven labels for training and verification, many proposed methods, which have achieved prominent cancer prediction results on unsure data, should be further reexamined in clinical practice. Meanwhile, such disenchantment should be set and reinforced in our community.

Moreover, there also exists another trade-off between learning from nodule regions and background. As demonstrated in \autoref{subsec::parameter} and \autoref{subsec::modalities}, sure data could achieve better performance if we regularize the attention bias to nodule regions. 
We do not deny the value of contextual information for nodule discrimination \citep{liu2021net}. Instead, we believe that for small training samples, concentrating more attention on nodule regions could help to learn domain-invariant features while reducing over-fitting, because background variables are more likely to be a confounding factor in such situation.
In fact, due to the difficulty in data acquisition, this task is hard to have massive sure data.


\begin{figure}[!hbt]
\centering
\includegraphics[width = 88 mm]{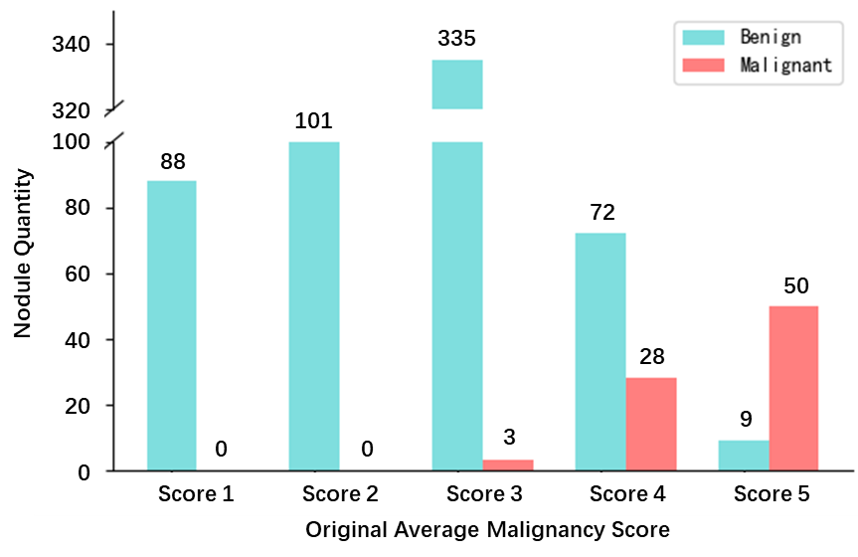}
\caption{Statistical result of LIDC-IDRI re-labeling nodules (benign or malignant) in terms of original average malignancy scores.
}
\label{fig8} 
\end{figure}

\subsection{Re-labeling LIDC-IDRI}



We use the new nodule diagnosis method in \autoref{subsec::nodule_dignosis} for LIDC-IDRI nodule relabel.
As shown in \autoref{fig8}, our re-labeled results are in broad agreement with the low original malignancy score ones. In score 3, the majority of the nodules are re-labeled to benign class, which could explain the better performance when the nodules of score 3 are assigned to the benign label in Scenario 123/45 in \autoref{sec::discussion_4}.
The new labels correct more than half of the original nodule labels with score 4 which could be the chief criminal leading to the data bias trouble because of the small suspicion on score 5 nodules.
Moreover, due to the lack of pathological ground truth, the relabel outcomes
of this study should always remain suspect until the LIDC-IDRI clinical information is available.

\subsection{Does the malignancy feature exists in CTs?}
\label{sec::discussion_7}


We finally discuss an interesting phenomenon in this subsection: with the similar number of training samples (scenario 12/45 in \autoref{sec::discussion_4}), model trained and tested using the only unsure data show much better classification performance than using only sure data.
The reasons are analyzed as follows:


\noindent\textbf{Relative performance for unsure data:}

Due to a lack of suitable reference to assess the likelihood of malignancy, low-level visible features (e.g. size, shape, brightness) are likely to be regarded as scoring criteria by radiologists’ observation. 
Built on consensus agreement within multiple radiologists, apparent features of these nodules can be easily extracted and classified by a commonly used model, whose power can successfully emulate the radiologist’s one. 

\autoref{fig7} also indicates that unsure data nodules with the same average scores often share similar features, whereas sure data presents more heterogeneous characteristics for intra-class nodules.
Thus, \cite{lei2020shape} is an effective scheme for LIDC-IDRI nodule classification by focusing on fine-grained features such as nodule shape and margins as its attention mechanism mimics radiologists’ reasoning.
In our method, we encode the fine-grained features into our model through SegNet and FNet.
However, without the sure data, this model will always take the human ability as a golden standard rather than real malignancy labels. This motivated us to build the sure dataset.


\noindent\textbf{Relative performance of sure data:}

According to the good performance on unsure data when using the same classifier, we can rule out the hypothesis that the model may have limited discrimination capability for lung nodule heterogeneity modeling.
However, no matter what input modality is used in \autoref{subsec::modalities}, the model performance using sure data cannot come close to that using unsure data.
Thus, we suspect that the 
image-based manifestation of pathological diagnosis may be unavailable in CT scans and give the argument as follow: 1) From the aspect of imaging principles, CT may contain limited relevant
information, such that the CT-based representation may lack high learnable correlation to its final diagnosis identified through a more sophisticated examination. 
This is analogous to ultraviolet rays that are invisible to humans. 2) From the aspect of nodule benign-malignant labels, it may not be reasonable if we subsume them under just two broad categories because the guideline for the definition of nodule malignancy is still inconclusive such as Adenocarcinoma in situ (AIS). Thus, we suppose it would be more appropriate to first classify the major types of nodules according to their visibility in CT volumes and then sort out the results to benign-malignant categories, if the nodule data is sufficient.


\section{Conclusion and future work}
In summary, we raised the vital issues that are commonly ignored in lung cancer prediction task from the aspect of unsure data and unreliable model reasoning.
For better verifiability of nodule diagnosis algorithm and authenticity that simulates the real clinical world, we constructed a sure dataset with pathologically-confirmed labels.
A synergic model was first proposed to integrate unsure data based on its properties and ultimately boost the classification performance of sure data.
Then, our ad-CSL loss treats the CAM not only a post-hoc interpretation to analyze a nodule classification process, but also as a participant to modify the classification process in such a way that the model could pay more attention to the faithful nodule features and gain improved generalizability. 
Moreover, similar nodule retrieval empowers a CAD system more practical for clinical application during collaboration with clinicians.
It is obvious that discriminating nodules of sure data is more difficult because they contain more complicated heterogeneity which is hard to model and there remains a critical dispute on CT-based manifestation of pathological diagnosis.
Besides, data scarcity still makes current nodule research a great challenge.
In the future, it is imperative to enable the deep learning system to possess the capacity of causal inference for lung cancer prediction. We will explore the approaches for explainable decisions that lead to a more creditable and faithful diagnosis.

\section*{Acknowledgments}
This work was partly supported by Shanghai Sailing Program (20YF1420800), National Nature Science Foundation of China (No.62003208), Shanghai Municipal of Science and Technology Project (Grant No. 20JC1419500), and Science and Technology Commission of Shanghai Municipality (Grant 20DZ2220400).

\bibliographystyle{model2-names.bst}\biboptions{authoryear}
\bibliography{paper}

\end{document}